\documentclass[10pt,preprint,eqsecnum]{aastex}
\usepackage{amsmath}

\begin{document}

\title{Star Formation and Feedback in Dwarf Galaxies}

\author{Shawfeng Dong\altaffilmark{1}, D. N. C. Lin\altaffilmark{1}
and S. D. Murray\altaffilmark{2}}

\altaffiltext{1}{UCO/Lick Observatory, University of California, Santa Cruz,
CA, 95064\\Email: dong@ucolick.org, lin@ucolick.org}
\altaffiltext{2}{University of California,
Lawrence Livermore National Laboratory, P.O. Box 808, Livermore,
CA 94550\\Email: sdmurray@llnl.gov}

\begin{abstract}
We examine the star formation history and stellar feedback effects of
dwarf galaxies under the influence of extragalactic
ultraviolet radiation.  Previous work has indicated that the
background UV flux can easily ionize the gas within typical dwarf
galaxies, delaying or even preventing cooling and star formation
within them.  Many dwarf galaxies within the Local Group are, however,
observed to contain multiple generations of stars, the oldest of which
formed in the early epochs of cosmic evolution, when the
background UV flux was intense.  In order to address this paradox, we
consider the dynamical evolution of gas in dwarf galaxies using a
one-dimensional, spherically symmetric, Lagrangian numerical scheme to
compute the effects of radiative transfer and photoionization.  We
include a physically-motivated star formation recipe and consider the
effects of feedback.
This scheme allows us to follow the history of the gas and of star
formation within dwarf galaxies, as influenced by both external and
internal UV radiation.  Our results indicate that star formation in the
severe environment of dwarf galaxies is a difficult and inefficient
process.  In potentials with total mass less than a few $10^6 M_\odot$,
and velocity dispersion less than a few km s$^{-1}$, residual gas is
efficiently photoionized by cosmic background UV radiation.  Since the
density scale height of the gas within these galaxies is comparable to
their size, gas may be tidally removed from them, leaving behind
starless residual dark matter clumps.  For intermediate mass systems,
such as the dSphs around the Galaxy, star formation can proceed with
in early cosmic epochs despite the intense background UV flux.
Triggering processes such as merger events, collisions, and tidal
disturbance can lead to density enhancements, reducing the recombination
timescale, allowing gas to cool and star formation to
proceed.  However, the star formation and gas retention efficiency may
vary widely in galaxies with similar dark matter potentials, because
they depend on many factors, such as the baryonic fraction, external
perturbation, IMF, and background UV intensity.  We suggest that the
presence of very old stars in these dwarf galaxies indicates that their
initial baryonic to dark matter content was comparable to the cosmic
value.  This constraint suggests that the initial density fluctuation of
baryonic matter may be correlated with that of the dark matter.  For the
more massive dwarf elliptical galaxies, the star formation efficiency
and gas retention rate is much higher.  Their mass to light ratio is
regulated by star formation feedback, and is expected to be nearly
independent of their absolute luminosity.  The results of our
theoretical models reproduce the observed $M/L-M_v$ correlation.
\end{abstract}


\section{Introduction}

According to the popular cosmological models dominated by cold dark
matter (CDM), the formation of large galaxies occurs in a hierarchical
fashion, through the coalescence of smaller systems (Blumenthal et
al. 1984; Navarro, Frenk, \& White 1997; Klypin, Nolthenius, \&
Primack 1997).  During mergers, the dissipationless dark matter
components combine to form the extended dark halos of larger galaxies.
The gas of the dwarf systems, however, dissipates its kinetic energy
during mergers, decouples from the dark matter, and falls either into
the center or onto the equatorial plane, forming the bulge and disk.
Dwarf galaxies observed today, such as the dwarf spheroidal (dSph)
satellites of the Milky Way, represent surviving members of the
original population of building blocks from which the Milky Way was
formed.

Along with the success of this picture, however, has come the
realization of new problems, both in the theory and in comparisons
with observations.  Three problems are of particular interest here.
One difficulty is the so-called ``over-cooling problem.''  In the CDM
scenario, the magnitude of linear density fluctuations,
$\sigma(M)\equiv \langle(\delta M / M)^2\rangle^{1/2}$ is a decreasing
function of the mass scale, $M$.  Such a fluctuation spectrum causes
low mass systems to become nonlinear and virialize at earlier epochs,
and within a more dense background, than larger systems.  Cooling is
very efficient at early epochs, due both to the generally higher
densities at high red redshift as well as the rapid increase in the
rate of inverse Compton cooling with redshift.  The early growth of
smaller systems, coupled with the more efficient cooling at early
epochs, leads to the expectation of rapid condensation of small, dense
gas clouds within the dark-matter halos of dwarf galaxies.  But,
unless star formation can be suppressed, most of the gaseous ordinary
matter would be converted into stars well before these galactic
building blocks can be assembled into present-day large galaxies such
as the Milky Way (White \& Rees 1978).

A second problem is the deficit of observed satellite dwarf galaxies
relative to theoretical expectations.  Observationally, the CDM
scenario naturally leads to a ``bottom-up'' hierarchical galaxy
formation picture that is qualitatively consistent with dwarf galaxies
having halos with higher phase-space densities than those around
larger normal galaxies.  High-resolution numerical simulations show,
however, that dwarf galaxies with higher internal densities retain the
integrity of their cores while their loosely bound halos merge with
each other during the formation of larger galaxies (Moore et al. 1999;
Klypin et al. 1999).  Consequently, the number density of surviving
satellite dwarf galaxies expected from the CDM scenario is much larger
than that observed in the Local Group.  Early bursts of star formation
could lead to the preservation of ordinary matter in a gaseous form,
followed by its removal from the dark matter potential.  Such a
process may account for the deficit of detectable satellite galaxies.

The third problem involves the building up of large galaxies from the
dwarf systems.  Under the action of dynamical friction from the
surrounding sea of dark matter, the dwarf galaxy building blocks lose
energy and orbital angular momentum before they eventually merge into
larger disk galaxies.  If the gas within the dwarf systems cools and
contracts within their dark matter potentials, the mergers are
delayed, leading to the formation of disk systems which are much
smaller and denser than observed (e.g., Navarro \& Benz 1991; Navarro
\& White 1994). The separation of gas from the dark matter potential
of dwarf galaxies is one possible mechanism for the ordinary matter to
retain sufficient angular momentum to form normal-size disk galaxies
(Weil, Eke \& Efstathiou 1998; Navarro \& Steinmetz 1997;
Sommer-Larsen, Gelato \& Vedel 1999).
 
Resolution of the above problems may require that the conversion of
gas into stars within the dwarf systems be inefficient.  Recent
theoretical investigations suggest that the conversion of gas into
stars in dwarf galaxies may be suppressed by the background UV
radiation at large redshifts (Weinberg, Hernquist \& Katz 1997;
Kepner, Babul \& Spergel 1997).  Integrating the full equations of
radiative transfer, heating, cooling and non-equilibrium chemistry for
nine species: $\mathrm{H}$, $\mathrm{H}$, $\mathrm{H^+}$,
$\mathrm{H^-}$, $\mathrm{H_2}$, $\mathrm{H^+_2}$, $\mathrm{He}$,
$\mathrm{He^+}$, $\mathrm{He^{++}}$, $\mathrm{e^-}$, Kepner et
al. (1997) computed the quasi-hydrostatic equilibrium states of gas in
spherically symmetric dark matter potentials roughly corresponding to
those of dwarf galaxies, and found that a typical background UV
radiation field can easily delay cooling and collapse of gas in halos
corresponding to $1 \ \sigma$ CDM perturbations with circular
velocities less than $30 \ {\rm km \ s^{-1}}$.

Those results, however, introduce an inconsistency with other observed
properties of low-mass dwarf galaxies.  The color-magnitude diagrams
of most dSphs indicate that star formation began within them at an
early epoch.  Indeed, Ursa Minor and Draco, whose stellar velocity
dispersions are only a few $\rm{km \ s^{-1}}$, and whose total masses
are less than $10^6 M_{\sun}$, are composed primarily of very old
stars.

In some dSphs, such as Carina and Fornax, the coexistence of multiple
generations of young and intermediate-age stars (Grebel 1997; Grebel
\& Stetson 1998) provides evidence of protracted and recent star
formation episodes. This is again difficult to account for, because
little atomic or molecular gas is found in them today (Knapp, Kerr \&
Bowers 1978; Mould et al. 1990; Bowen et al 1997; Carignan et
al. 1998), though some gas has been detected around some more distant
dSphs, such as Sculptor. While it is natural to expect that either
internal star formation or external ram-pressure stripping might
easily remove any residual gas from the shallow potentials of dSphs
(Noriega-Crespo et al. 1989; Mori \& Burkert 2000), it is difficult to
reconcile the lack of either cool or warm gas with the presence of
young stars.

A more complete picture of the early evolution of dwarf galaxies, and
of their role in the formation of massive systems, therefore requires
a more thorough understanding of the various physical processes, both
external and internal, that act to trigger and regulate star formation
within them.  In this paper, we present an investigation of the
hydrodynamic evolution and star formation history of dwarf galaxies.
We construct a code which integrates a realistic star formation
recipe, feedback, and radiative transfer in a 1D spherically symmetric
Lagrangian scheme.  The code allows us to follow the evolution of both
the state of the gas and the star formation of dwarf galaxies
subjected to different background UV fluxes and different external
perturbations.  In examining the parameter space, we seek answers to
the following questions:

\noindent
1). Is star-formation in dwarf galaxies triggered by a decrease in the
background UV radiation, or by external perturbations?

\noindent
2). What is the star formation efficiency, {\it i.e.} the fraction of the
initial mass of gas that is converted into stars?

\noindent
3). What is the mechanism for triggering multiple bursts of star-formation,
such as are observed in some dSphs?

In \S2 of this paper, we describe the dwarf galaxy model, the star
formation recipe, feedback, and the Str\"omgren shell model for
radiative transfer and photoionization.  We combine all of these into
a 1D spherically symmetric Lagrangian hydrodynamical scheme. In \S3 we
discuss the results of our simulations, and in \S4 we discuss the
implications of our results.

\section{Numerical Models}

A wide range of physical processes and physical scales are associated
with galaxy and star formation.  Including all of them would be a
formidable task, beyond the capabilities of even the most powerful
current generation of computers.  In this work, we simplify the
problem by examining the time evolution of a spherically symmetric
cloud, and combine models for the most important processes, including
radiative transfer, photoionization, star formation, and feedback.
The problem therefore becomes tractable, and the results which we
obtain may be included in larger-scale simulations.
 
\subsection{Dark Matter Halo}

We simulate the evolution of the gas within a fixed halo potential.
Similar to the approach adopted by Kepner et al. (1997), we choose a
form for the potential found by Burkert (1995), which has been shown
to be a good fit to many dwarf systems.  The corresponding dark matter
density distribution is given by
\begin{equation}
\rho_{DM} = \frac{\rho_0 r_0^3}{(r + r_0) (r^2 + r_0^2)}
\end{equation}
where
\begin{equation}
\rho_0 = 4.5 \times 10^{-2} (r_0/ \rm{kpc})^{-2/3} M_{\sun} \rm{pc}^{-3}
\end{equation}
is the central density.  The dark matter density tends to
$\rho\propto const.$ for $r \ll r_0$, and $\rho\propto r^{-3}$ for
$r \gg r_0$.
The rotational velocity, and mass at $r_0$ are given by
\begin{equation}
v_0 = 17.7 (r_0 / \rm{kpc})^{2/3} \rm{km \ s^{-1}}
\end{equation}
\begin{equation}
M_0 = 7.2 \times 10^7 (r_0 / \rm{kpc})^{7/3} M_{\sun}
\end{equation}
In the numerical simulations presented below, the above mass
distribution is arbitrarily assumed to extend out to a halo radius
\begin{equation}
r_{halo} = 3.4 r_0,
\end{equation}
and so the total mass of the dark matter halo is
$M_{DM} = 4 \pi \int_{0}^{r_{halo}} \rho_{DM}
r^2 dr = 5.8 M_0$.

\subsection{Gaseous Component}

The dark matter halo contains ordinary matter in the forms of stars
and gas.  The evolution of the gas density, $\rho_g$ is described by
the continuity equation,
\begin{equation}
\frac{{\rm{d}} \rho_{g}}{{\rm{d}} t} + \rho_g \nabla \cdot \mathbf{v} =
  - \dot{\rho_s}.
\label{eq:continuity}
\end{equation}
The sink term $\dot \rho_s$, due to star formation, denotes the rate
by which the gas becomes gravitationally unstable (see
Equation~\ref{eq:rhos}).  In the above expression,
${\rm{d}}/{\rm{d}}t$ is the material derivative and $\mathbf{v}$ is
the gas velocity.

The momentum equation for the gas is
\begin{equation}
\rho_g \frac{{\rm{d}} \mathbf{v}}{{\rm{d}}t} = \rho_{g} \mathbf{g} +
\nabla p + \nabla p_k,
\label{eq:momentum}
\end{equation}
where the force $\mathbf{g} = -\nabla \psi$ is due to the gradient of
the gravitational potential $\psi$, $p$ is the gas pressure, and we
introduce a kinetic pressure, $p_k$ to represent the momentum input
due to stellar wind and supernovae of massive stars (see
Equation~\ref{eq:pk}).

Massive stars emit intense UV radiation, providing another mechanism
for feedback.  In
addition, observations of Ly$\alpha$ clouds along quasar
lines-of-sight suggest that, at high redshifts, the universe is
permeated by a metagalactic UV flux.  At $z=2$, this background is
estimated to have a strength $J_\nu \approx 10^{-21} \ \rm{ergs \
s^{-1} \ cm^{-2} \ ster^{-1} \ Hz^{-1}}$ at the Lyman limit (Bechtold
et al 1987).  Both of these energy sources are modelled in the
energy equation of the gas,
\begin{equation}
\frac{{\rm{d}} \epsilon}{{\rm{d}}t}+p\frac{{\rm{d}} V}{\rm{d}t}=G-L,
\label{eq:energy}
\end{equation}
where $\epsilon$ is the specific internal energy, $V=1/\rho_g$, and $G$ and
$L$ represent radiative heating and cooling terms.

\subsection{Prescription for Star Formation and Feedback}

In order to consider the effect of feedback, the formation rate of
massive stars must be specified.  Here, we introduce a simple
prescription for massive star formation. In general, gas is
destabilized by its collective self-gravity and stablized by internal
pressure and external tidal stress.  When the residual gas within a
galaxy is fully ionized, it attains hydrostatic equilibrium, with the
Jeans mass $M_J$ of the system comparable to or larger than the total
mass of gas.  But, if the gas is able to cool well below the
ionization temperature, $M_J$ decreases to masses comparable to high
mass stars. The density of the residual gas also increases as it
contracts to a new hydrostatic equilibrium within the dark matter
potential.  In those regions of the dwarf galaxy where $ \rho_g > 3
\langle \rho_{DM} (r) \rangle $, with
\begin{equation}
\langle\rho_{DM} (r)\rangle= (3/r^3) \int_0 ^r \rho_{DM}(r^\prime)
r^{\prime 2} d r^\prime,
\end{equation}
the tidal field of the potential is insufficient to prevent the
collapse of gas within the Jeans radius $R_J = \sqrt{\frac{3 R_g T}
{1.6 \pi G \rho_g \mu}}$.  In the absence of a strong magnetic field,
the gas collapses to form stellar mass objects when these conditions
are satisfied.

Ionizing photons and stellar winds from young, massive stars act to
raise the Jeans mass of surrounding gas, inhibiting further star
formation.  The collapse of the gas cloud does not, however, lead to
immediate star formation and feedback.  A minimum time delay between
the instant when the gas becomes gravitationally unstable and the
onset of main sequence evolution for the resulting generation of stars
is the dynamical free-fall time scale $\tau_{ff} = (G \rho_g)^{-1/2}$.
If the collapsing gas has angular momentum, protostellar disks would
form, through which gas is accreted onto the growing protostellar
objects on a viscous time scale.  After the formation of the
protostars, an additional time interval, $\tau_{onset}$, must elapse
before the emission of intense UV radiation. For most of our models,
we choose $\tau_{onset} = 0.5$~Myr.  But in some delayed star
formation models, we choose $\tau_{onset} = $ either 5 or 50~Myr, to
represent delays due to disk accretion (Yorke \& Sonnhalter 2002),
merger processes (Murray \& Lin 1996, Bonnel {\it et al.} 1998), or
sequential star formation (Elmegreen \& Lada 1977).  Thus, changes in
the stellar and gas mass densities ($\rho_{\star}$ and $\rho_g$
respectively) are related via a time-delay, such that
\begin{equation}
\dot{\rho_{\star}}(t) = \dot{\rho_s} (t - \tau_{ff} - \tau_{onset})
\end{equation}
and,
\begin{equation}
\dot{\rho_s} = \beta \frac{\rho_g}{\tau_{ff}}
\label{eq:rhos}
\end{equation}
where $\tau_{ff} $ refers to the local free-fall timescale at the epoch when
the collapse begins, i.e.  $t - \tau_{ff} - \tau_{onset}$.  We adopt a
form for the star formation efficiency, $\beta$, of
\begin{equation}
\beta = \left(\frac{\tau_{onset}}{\tau_{onset} + \tau_{ff}}\right)
\frac{\beta_{max}}
{\left[1+\left(\frac{T}{T_J}\right)^{\lambda_T}\right]
\left[1+\left(\frac{R_J}{R_R}\right)^{\lambda_R}\right]}
\end{equation}
where $T_J = 0.4 (\frac{4 \pi} {3}) ^{\frac{1}{3}} \mu G
M_s^{\frac{2}{3}} \rho_g^{\frac{1}{3}}$ is the critical temperature for
the onset of $M_s = 100 M_{\sun}$ star formation, and $R_R$ is the Roche 
radius of the gas in the potential of the dwarf galaxy. The three factors
in $\beta$ act to decrease the star formation efficiency in situations
where it is expected to be low.  When the local free fall time is
longer than the timescale for the onset of main sequence evolution,
then it would be expected that, because of the spread in star
formation times, the first stars to form would inhibit further star
formation.  The efficiency is reduced for high stellar masses, due to the
fact that the dynamical timescale for the evolution of the gas exceeds the
evolution time of massive stars, in which case the first stars to form
would re-ionize the surrounding gas before it, too, could collapse to form
stars.  Finally, star formation is also expected to be inefficient whenever
the Roche radius of a Jeans unstable region exceeds its tidal radius
within the dwarf galaxy. 
We set the star formation quenching factors $\lambda_T=\lambda_R=4$,
such that star formation is rapidly shut off whenever one or more of
the above criteria is met.  Unless otherwise stated, we set the
maximum efficiency $\beta_{max}=1$.  In some cases, we choose
$\beta_{max} = 0.1$, representing a conversion of at most ten percent
of the total residual gas into stars within a free fall time.

Accompanying each massive progenitor star, there are many coeval
lower-mass stars.  Thus, the corresponding formation rate, per unit
volume, of massive UV and wind producing stars is given by
\begin{equation}
\dot{n_{\star}} = \frac{\dot{\rho_{\star}}}{M_c}.
\end{equation}
The total mass of stars formed per UV, wind, and supernova producing
star, $M_c$, can be estimated from a Scalo initial mass function (IMF)
to be 150 $M_{\sun}$ (Scalo 1986; Miller \& Scalo 1979).  For more
general initial mass functions, we set
\begin{equation}
M_c = f_{IMF} 150 M_{\sun},
\end{equation}
where $f_{IMF}>1$ for a steeper initial mass function with a
relatively small cut off in the stellar mass function at the high end.
Unless otherwise stated, $f_{IMF} = 1$ is chosen for most of our
models.  In order to explore the effects of variations, we examine
some models in which $f_{IMF}\gg 1$.

The formation of massive stars strongly influences the environment.
The massive stars emit UV radiation with a well-determined
flux (see e.g. Osterbrock 1989).  When averaged over a Miller-Scalo
initial mass function, the rate of ionizing photon emission per
10~$M_{\sun}$ star is $\overline{Q}_\star = 10^{48.5}$ s$^{-1}$.
Photoionization by UV radiation from massive stars can be included in the
energy equation of the gas, and in determining its ionization state
(see below).  Stars are assumed to emit UV photons at an average rate of
$\overline{Q}_\star$ for a time equal to the average lifetime of a
massive star, $\tau_\star=4$~Myr.  After that time, they are taken to be
``dead stars,'' contributing to the overall mass of the system, but
providing no feedback to the gas.

After a brief stage of adiabatic expansion away
from the massive stars, the main effect of the stellar winds and
supernovae is to increase the momentum of the surrounding gas.  We
approximate this feedback process as a a ``kinetic pressure'' term in
Equation~(\ref{eq:momentum}), defined for each given volume (zone) of
the model as
\begin{equation}
p_k  = \frac{d \dot{N}_{sn}}{d S} P_{sn} +
        \frac{d N_{\star}}{d S} \dot{m}_w v_w,
\label{eq:pk}
\end{equation}
where the surface area at the center of the zone is $S$.  The first
term in Equation~\ref{eq:pk} is the momentum flux due to supernovae,
and the second is that due to stellar winds.  In the above equation,
$P_{sn}$ is the momentum of a single supernova explosion (we adopt
$4.5 \times 10^{42}$g cm s$^{-1}$), $\dot{N_{sn}}$ is the total
supernova rate (number of SN per unit time) within the volume of the
zone.  The value of $\dot{N_{sn}}$ is obtained from the the supernova
rate per unit volume, which itself is determined by the star formation
rate at the earlier time $t-\tau_{\star}$, i.e.
\begin{equation}
\dot{n}_{sn}(t) = \dot{n}_{\star} (t - \tau_{\star}).
\end{equation}
The total number of massive stars on the main sequence within the volume
of the zone is $N_{\star}$, while the wind mass loss rate of a typical
massive star and the typical wind speed are $\dot{m}_w (= 10^{-6}
M_{\sun}$ y$^{-1}$) and $v_m (= 10^3$ km s$^{-1}$), respectively.

With these prescriptions for the feedback process, we use a
one-dimensional Lagrangian hydrodynamic model for the evolution of the
baryonic gas component within the fixed external gravitational
potential set by the dark matter (Richtmyer \& Morton 1995).

\subsection{Photoionization and Radiative Transfer}

As indicated above, UV radiation from both massive stars and the
background can lead to photoionization and heating of the residual
gas.  A comprehensive treatment of the full 3D radiative transfer
requires the determination of the contribution to the flux at every
point from every other point along all paths for each wavelength,
which is at the minimum a 6D problem. However, in most instances,
symmetries can be introduced, resulting in a more tractable situation.
The greatest simplification occurs when the gas can be assumed to be
optically thin throughout.  This approximation is sufficient in the
majority of cosmological situations (Katz, Weinberg \& Hernquist 1996;
Navarro \& Steinmetz 1996; Anninos, Norman \& Clarke 1994) and only
breaks down in the cores of halos that have undergone sufficient
cooling, a situation that is usually made intractable by the
complexities of star formation.

The next simplest geometry is that of a slab (or a sphere under the
assumption of a radially perpendicular radiation field), which reduces
the flow to an intrinsically 2D problem.  This approach is the most
common in radiative transfer, and has been used to address similar
situations to those considered here (Hainman, Rees \& Loeb 1996;
Haiman, Thoul, \& Loeb 1996).

\begin{figure}
\plotone{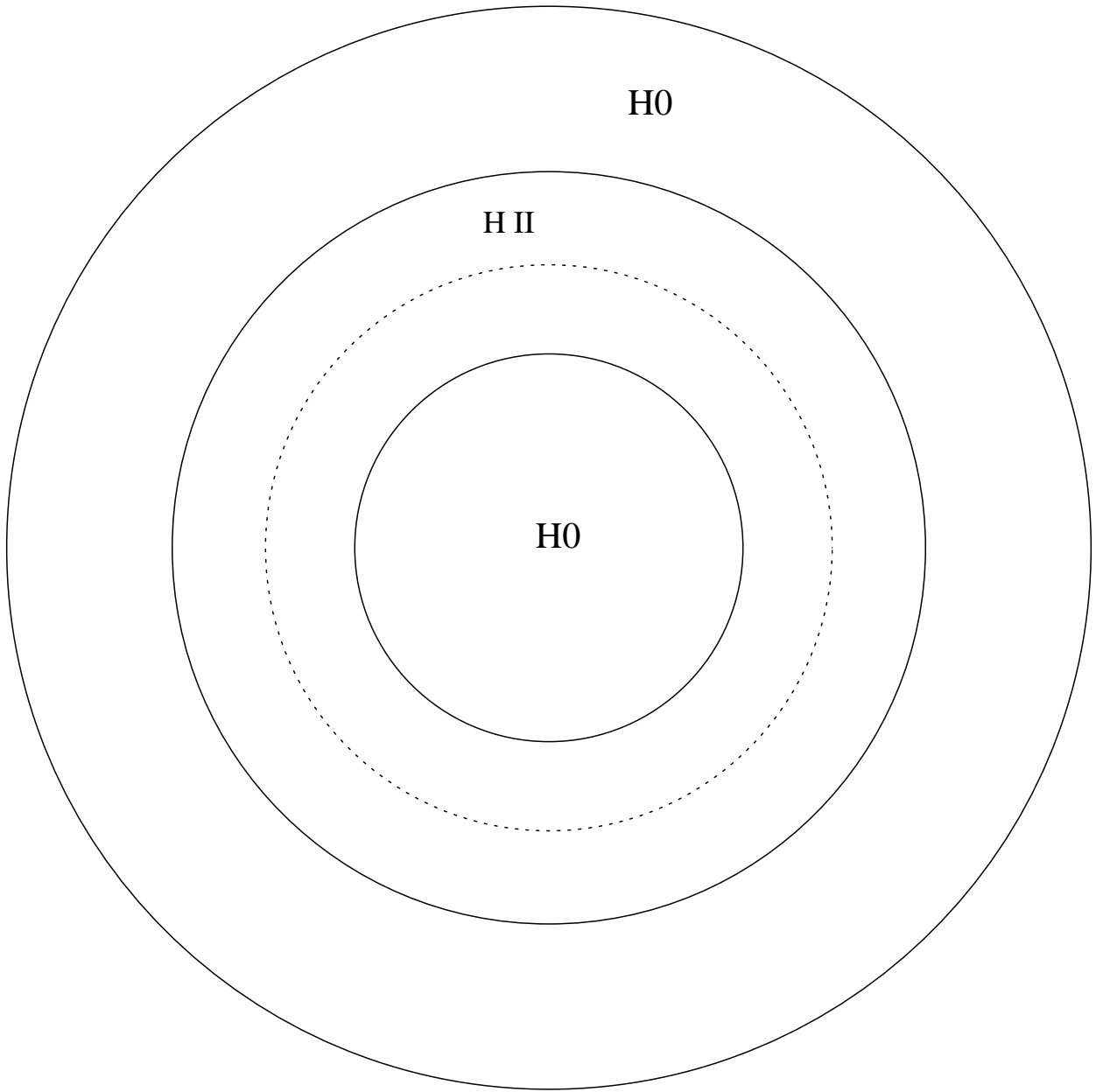}
\caption{Str\"omgren shell model. }
\label{fig:strom}
\end{figure}

In this paper, we adopt a similar approach for spherical geometry.  We
relax the assumption of a radially perpendicular radiation field by
treating radiative transfer using the Str\"omgren shell model.
Similar to the original Str\"omgrem model, the ionized ``H II'' shell
is assumed to be surrounded isotropically by a neutral $\mathrm{H}^0$
region (Figure~\ref{fig:strom}). The rate of entry of ionizing photons
into the ionized volume balances the total number of recombinations to
excited levels within the volume.

\begin{figure}
\plotone{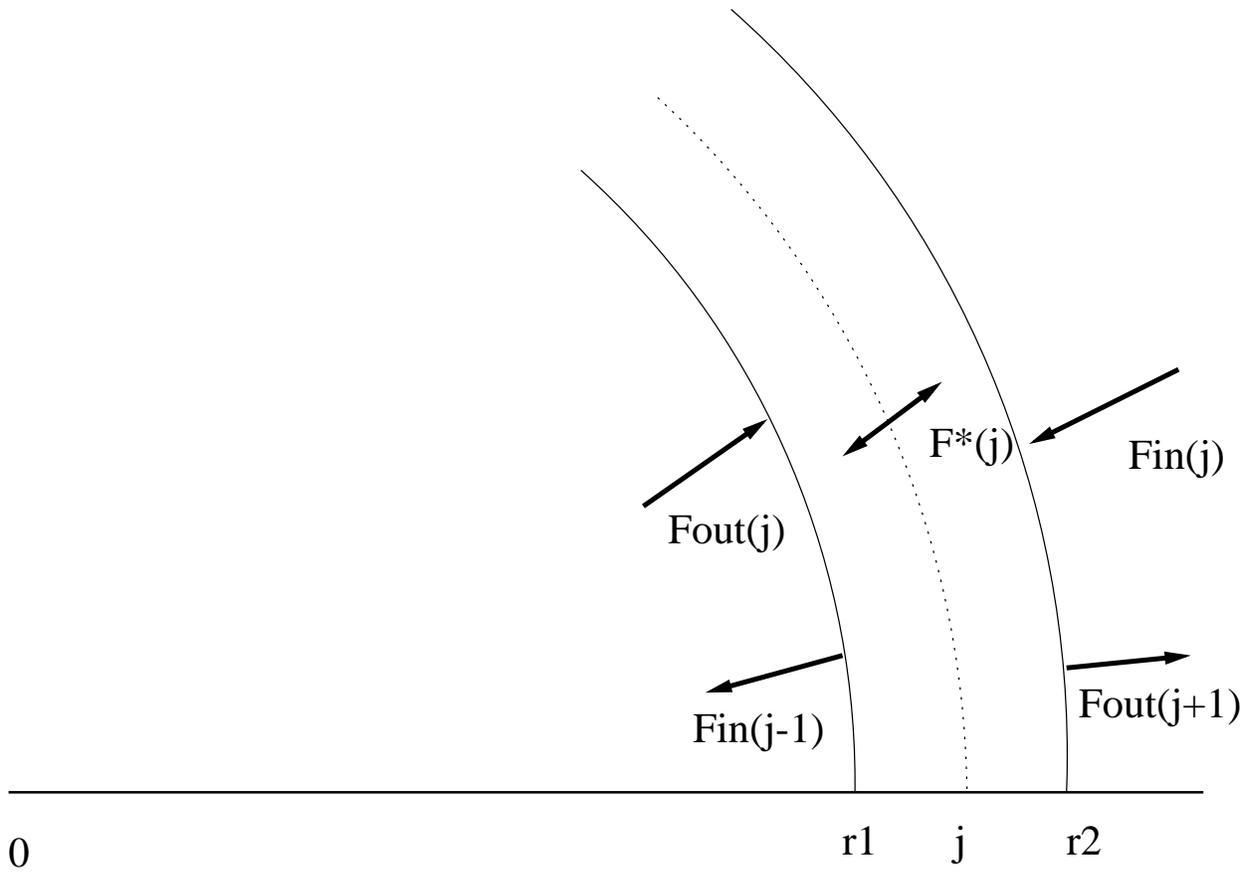}
\caption{Radiative transfer at zone j. }
\label{fig:geom}
\end{figure}

In our implementation of the Str\"omgrem shell model, we use a
discretized 1D Lagrangian hydrodynamic scheme.  We define
\begin{equation}
Q_j = V_j n_H^2 \alpha_B
\end{equation}
to be the number of photons required to ionize zone $j$, with $V_j$
being the volume of the zone, $n_H$ being the number density of
hydrogen, and $\alpha_B$ being the case B recombination coefficient
(Osterbrock 1989).  The total number of UV photons emitted by the
massive stars in the zone is $F_{\star,j}$, while $F_{in,j}$ is the
rate of UV photons entering into the zone from larger radii, and
$F_{out,j}$ is the rate of UV photons entering into the zone from
smaller radii (Figure~\ref{fig:geom}). In the innermost zone, there is
only $F_{\star}$ and $F_{in}$.

The total rate of ionizing photons either produced within or entering
into the zone is $F_{t,j} = F_{\star,j} + F_{in,j} + F_{out,j}$.  If
$F_{t,j} < Q_j$, then the zone is only partially ionized, and the
ionization fraction of the gas is given by
\begin{equation}
X_j = F_{t,j} / Q_j.
\end{equation}
If $F_{t,j}>Q_j,$ then $X_j = 1$, and any photons in excess of $Q_j$
give rise to $F_{out,j+1}$ and $F_{in,j-1}$ of adjacent zones, as
follows;
\begin{equation}
F_{in,j-1} = f_g \times \left(\frac{1}{2}F_{\star,j} + F_{in,j} -
\frac{1}{2}Q_j \right)
\end{equation}
and
\begin{equation}
F_{out,j+1} = \frac{1}{2}F_{\star,j} + F_{out,j} - \frac{1}{2}Q_j +
(1 - f_g) \times \left(\frac{1}{2}F_{\star,j} + F_{in,j} -
\frac{1}{2}Q_j\right).
\end{equation}
The term
\begin{equation}
f_g = (r_{in}/r_{out})^2
\end{equation}
with $r_{in}$ and $r_{out}$ being, respectively, the inner and outer
radii of any zone, is a factor that takes into consideration the
spherical geometry of the model.  In the limit of isotropic radiation,
that geometry causes slightly more than half of the photons within a
zone to exit in the outward direction.  The case of either slab
geometry or of a radially-directed radiation field can be represented
by $f_g=1$.  The above form of $f_g$ can be arrived at by considering
the optically thin limit with only inflowing UV photons.  In that
case, the UV intensity $ J_{\nu}(r_{in}) \approx J_{\nu}(r_{out})$,
and the flux of photons $F \propto r^2$, thus $F(r_{in}) / F(r_{out})
\approx (r_{in}/r_{out})^2$.  Care is also taken to deal with the
special cases when either $F_{in,j-1}$ or $F_{out,j+1}$ is negative.
Iterations by sweeping inwards and then outwards several times are
taken to ensure convergence.  For the innermost zone, $F_{out}=0$,
while for the outermost zone $F_{in}$ is given by the external
radiation field.

The external UV flux $J_{\nu}\approx
10^{-23}$~ergs~s$^{-1}$~cm$^{-2}$~ster$^{-1}$~Hz$^{-1}$ at the Lyman
limit at low red shift, and $50$ times more intense at redshift $z=3$
(cf. Haardt \& Madau 1996), is sufficient to ionize most dwarf
spheroidal galaxies throughout.  For a less intense external flux,
neutral residual gas is surrounded by ionized gas.  The presence of
metals would increase the critical $J_{\nu}$ needed for fully ionizing
the dwarf galaxy, but, because we are considering the early evolution
of low-metallicity systems, the modification is expected to be
limited.  The temperature of the photoionized gas is taken to be
15,000~K, appropriate for low-metallicities.  We calculate the
temperature for the $\mathrm{H}^0$ region according to
Equation~\ref{eq:energy}.  We adopt equilibrium cooling rates,
including contributions from hydrogen, helium, and metals (Dalgarno \&
McCray 1972; Fall \& Rees 1985).  Because the dwarf galaxies being
modeled have low metallicities, and we are modeling them at early
epochs, the metallicity is taken to be low, [Fe/H]=-2.  In the absence
of photoionization, such metal-poor gas gas may cool to a minimum
temperature of $\sim 10^2$~K (Kang et al. 1990; Lin \& Murray 1992).
Using the above treatments for heating, cooling, and radiative
transfer, we can compute the ionization fraction and the temperature
of the gas as well as the attenuation of the ionizing photons.

We note that the model described above follows the assumptions used in
calculating static, ``initial'' Str\"omgren spheres.  These are
formed extremely rapidly, on timescales of the
order of the recombination time of the gas ($\sim10^8$~s).  Subsequently,
the gas expands in response to the photoionization heating.
To compute, on galactic scales, the evolution of individual Str\"omgren
spheres, as well as that of the ionization front propagating inwards from
the external UV radiation, is a multidimensional problem, beyond current
computational capabilities. Our models do capture the expansion of the
ionized gas
within each zone in response to heating, and therefore account for its
global effects upon feedback and star formation.  The essential physics
of the problem is therefore included in the models.

Our models also assume the gas to be uniformly distributed within each
zone.  Clumpiness in 
the ISM would increase the value of $Q_j$ from that given above.  Clumpiness
would result from either temperature variations (a multi-phase medium) or
self-gravity.  In the former case, the presence of a cold medium implies the
lack of sufficient UV photons to completely ionize the galaxy, and that
subsequent star formation shall rapidly lead to re-heating and ionization
of the gas, ie. in our models, this corresponds to a dwarf galaxy undergoing
star formation.  If the gas is completely heated, the Jeans mass is
sufficiently large that self-gravity shall be unable to cause significant
clumping.

Finally, our models do not include metallicity evolution of the gas, due
to star formation.  Metallicity does not strongly affect the size of
Str\"omgren spheres, and so this shall not significantly affect our 
assumptions regarding ionization by external and internal UV sources.
The primary affect of increasing metallicity is to enhance cooling in gas
that has lost its heat source, potentially enhancing the star formation
efficiency.  This, too, is not likely to be a strong effect, because the
timescales for star formation are dominated by the initial free-fall time
of the gas, and the pre-main sequence evolution time of the stars.  The
main affect would, therefore, likely be in the stellar mass function.  
Higher metallicity gas can cool to lower temperatures, possibly enhancing
the formation of low mass stars, and reducing the effectiveness of internal
feedback.  Such effects must be included in the uncertainties of our models,
and we explore a wide range of parameters in order to examine their
potential significance.

\section{Dwarf Galaxy Models}

\subsection{Model Parameters}

The currently preferred (e.g. de Bernardis et al 2002) cosmological
parameters are as follows: the Hubble constant is
\begin{equation}
H_0 = 100 \:h \ \rm{km~s^{-1} \ Mpc^{-1}} =
71 \pm 3.5 \ \rm{km~s^{-1} \ Mpc};
\end{equation}
the age of the universe is $t_0 = 13.7 \pm 0.1$~Gyr; the baryonic
density parameter is $\Omega_b = 0.04$; and the dark matter density
parameter is $\Omega_m = 0.23$.

These parameters are relevant to the issues of delayed star formation
in dwarf galaxies.  For instance, Kepner et al. (1997) simulated halos
in the range of $ 15 \ \rm{km \ s^{-1}} < v_c < 70 \ \rm{km \ s^{-1}}
$, which, using their model, translate to: $3 \times 10^8 M_{\sun} <
M_{halo} < 5 \times 10^{10} M_{\sun}$.  In their models, 
$3 \times 10^5 M_{\sun} <
M_{gas} < 5 \times 10^9 M_{\sun}$, such that the baryonic fraction is:
\begin{equation}
10^{-3} < b \equiv M_{gas} / M_{halo} < 0.1.
\end{equation}
With the very small value of $b=0.001$, dwarf
galaxies at the low end of the mass range could not form stars even
today because the recombination efficiency is insufficient to offset
photoionization due to external background UV radiation.  Kepner et
al. (1997) deduced such a small value for $b$ under the assumption
that the ordinary matter is accreted onto the dark matter halo from an
ionized intergalactic medium with temperature $T=2\times10^4$~K.
However, if the dwarf galaxies formed prior to the era of reionization
or if the first generation stars are primarily formed in the
progenitors of dwarf galaxies, then cool, neutral gas may be retained
by their dark matter haloes with the cosmological value of
$b=\Omega_b/\Omega_m \sim 0.17$.

Observation of dwarf galaxies in the local group find high
baryonic fractions (Mateo 1998).  Indeed, in some systems, such as
NGC~147, the observationally inferred estimate of $b$ approaches
unity.  Such values indicate that these systems may have undergone
complicated and highly varied evolutionary histories.  It is possible
that dwarf galaxies had larger values of $b$ in the past, and have
since preferentially lost a fraction of their baryonic matter through
star formation feedback (Dekel \& Silk 1986) and tidal stripping
(Murray, Dong, \& Lin 2003).  It is also possible that the observed
dwarf galaxies are completely different from the dwarf galaxy building
blocks in the CDM scenario.  In view of these uncertainties, it is
useful to consider a wider range of parameters than adopted by Kepner
et al. (1997).  We find below that the magnitude of $b$ is a crucial
factor in regulating the star formation efficiency in dwarf galaxies.

In the analysis of Kepner et al. (1997), gas is assumed to be in
quasi-hydrostatic equilibrium at all times.  Yet, both pre-star
formation evolution and post-star formation feedback processes lead to
flow and evolution of the residual gas mass that should be treated
with a time-dependent dynamical calculation.  In addition, external
perturbations, such as tidal disturbances by nearby large galaxies or
collisions with other gas-rich dwarf galaxies are also important in
modifying the density and velocity of the gas.  For some models, we
approximate the possibility of a strong tidal compression (or a mild
collision) with an initial velocity perturbation in the form of a
Hubble inflow
\begin{equation}
v_i = - H_p r
\label{eqn:compeqn}
\end{equation}
When scaled with the halo radius and the characteristic velocity
dispersion, the above expression is equivalent to $v_i/v_0 = - h_p (r
/ r_{halo})$.

The main purpose of our calculation is to study the conditions under
which old stars formed within dwarf galaxies.  The age of these stars,
10-12~Gyr, corresponds to high-red-shift epochs, when the external UV
flux is intense.  We approximate the frequency dependence of the
external UV energy flux by
\begin{equation}
J_{\nu}(z) = J_0(z) (\nu / \nu_0)^{- \alpha}.
\end{equation}
For the isotropic external flux, the rate at which photons enter the
dwarf galaxy ($\rm{photons/s}$), whose outer surface area is $S$, is
\begin{equation}
Q_{ext} = S \times \int_{\nu_0}^{\infty}
          \frac{2 \pi J_{\nu}}{h \nu} d \nu
         =  \frac{2 \pi J_0 S}{h \alpha}.
\end{equation}
The flux of photons across the surface would involve integrating over
$\mu{\ \rm d}\mu$ where $\mu={\rm cos}{\ }\theta$, and $\theta$ is the
angle from the normal.  The resulting value of $Q_{ext}$ would be smaller
by a factor of two.  This introduces some uncertainty into our results,
but is smaller than the uncertainty in the magnitude of the UV
background, and much smaller than the range in $Q_{ext}$ experienced
by dwarf galaxies over their lifetimes.
We adopt $\alpha = 1.5$ in all our calculations. 

Ionizing photons are also generated by the massive stars formed within
these galaxies.  The UV production rate by massive stars in each zone
($\rm{photons/s}$) is
\begin{equation}
Q_{\star j} = N_{\star j} \overline{Q}_\star,
\end{equation}
where $N_{\star j}$ is the number of massive stars of zone, and
$\overline{Q}_\star$ is the value of $Q_\star$ averaged over massive
stars according to the Miller-Scalo IMF ($\overline{Q}_\star =
10^{48.5}$ s$^{-1}$).

\subsection{The Basic Models: Effect of Baryonic Fraction}

\subsubsection{Dwarf Galaxies with Modest Velocity Dispersion}

The parameters of the models that we have examined are shown in
Table~\ref{tab:galaxymodels1}.  In the table are listed, for each
model, the scale radius ($r_0$), the initial baryonic-to-dark matter
ratio ($b$), the ratio of the background flux to its value at the current
epoch, the upper limit upon the star formation efficiency
($\beta_{max}$), the falue of $f_{IMF}$, the value of $\tau_{onset}$,
and the fraction of gas converted into stars, $f_\star$.

In the first series of three models, A1, A2, and A3, we simulate the
evolution of an unperturbed dwarf galaxy ($H_p =0$) at the present
epoch ($J_0 = 4. \times 10^{-23} \rm{erg \ s^{-1} \ cm^{-2} \
ster^{-1} \ Hz^{-1}}$).  The scale length of the potential $r_0 =
0.6$~kpc, such that $M_0 = 2.2 \times 10^7$ $M_{\sun}$, $v_0=12.6
\rm\ {km \ s^{-1}}$, and $M_{DM} = 1.3 \times 10^8 M_{\sun}$.  These
model parameters for the halo strcuture are comparable to those of
Leo I and other dSph (Mateo {\it et al.} 1998).  In order
to demonstrate the dependence of the star formation efficiency on the
ordinary matter content, we consider $b=0.001, 0.1, $ and $ 0.2$.  In
these and all other models, the ordinary matter has the same initial
spatial density distribution as the dark matter ($\rho_{g,i} = b \,
\rho_{DM}$), and an initial temperature of $10$~K everywhere.

\begin{figure}
\plotone{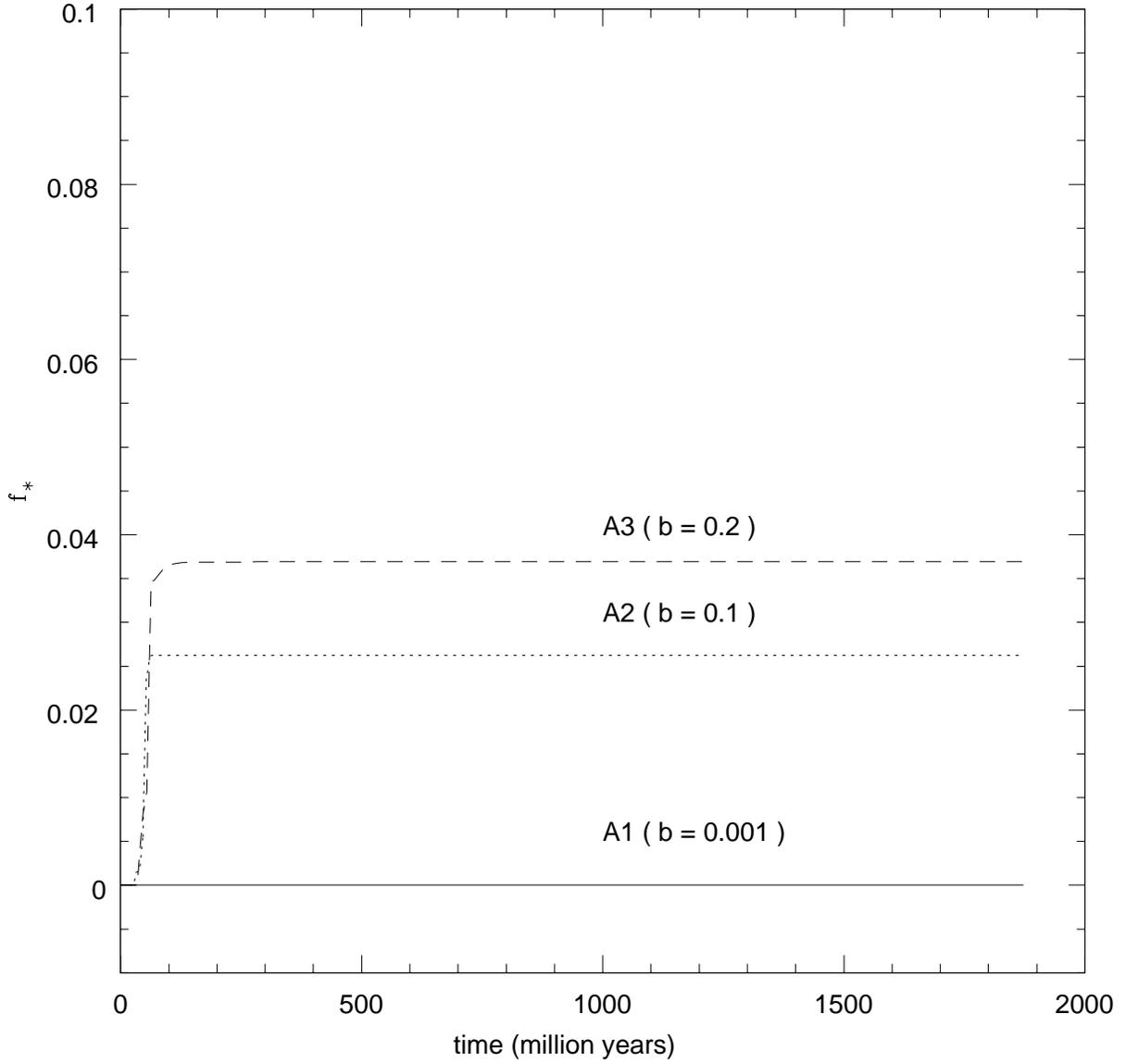}
\caption{Star formation history for models with $r_0 = 0.6$~kpc, $J_0
= 4 \times 10^{-23} $, and $b = 0.001, 0.1$ and $0.2$ respectively for
model A1, A2 and A3.}
\label{fig:r006}
\end{figure}

In Figure~\ref{fig:r006}, we plot the fraction, $f_\star$, of ordinary
mass that is converted into stars as a function of time.  For the
tenuous ordinary matter distribution in model A1, the residual H is
fully ionized, and no stars are formed.  Despite the low initial
temperature, the residual gas is ionized and heated to $T=1.5 \times
10^4$~K almost instantly. As the gas evolves toward a new equilibrium,
most of it leaves the galaxy.  For the residual gas, the recombination
process is inefficient compared with the photoionization even for the
present low flux of background ionizing photons.

\begin{figure}
\plotone{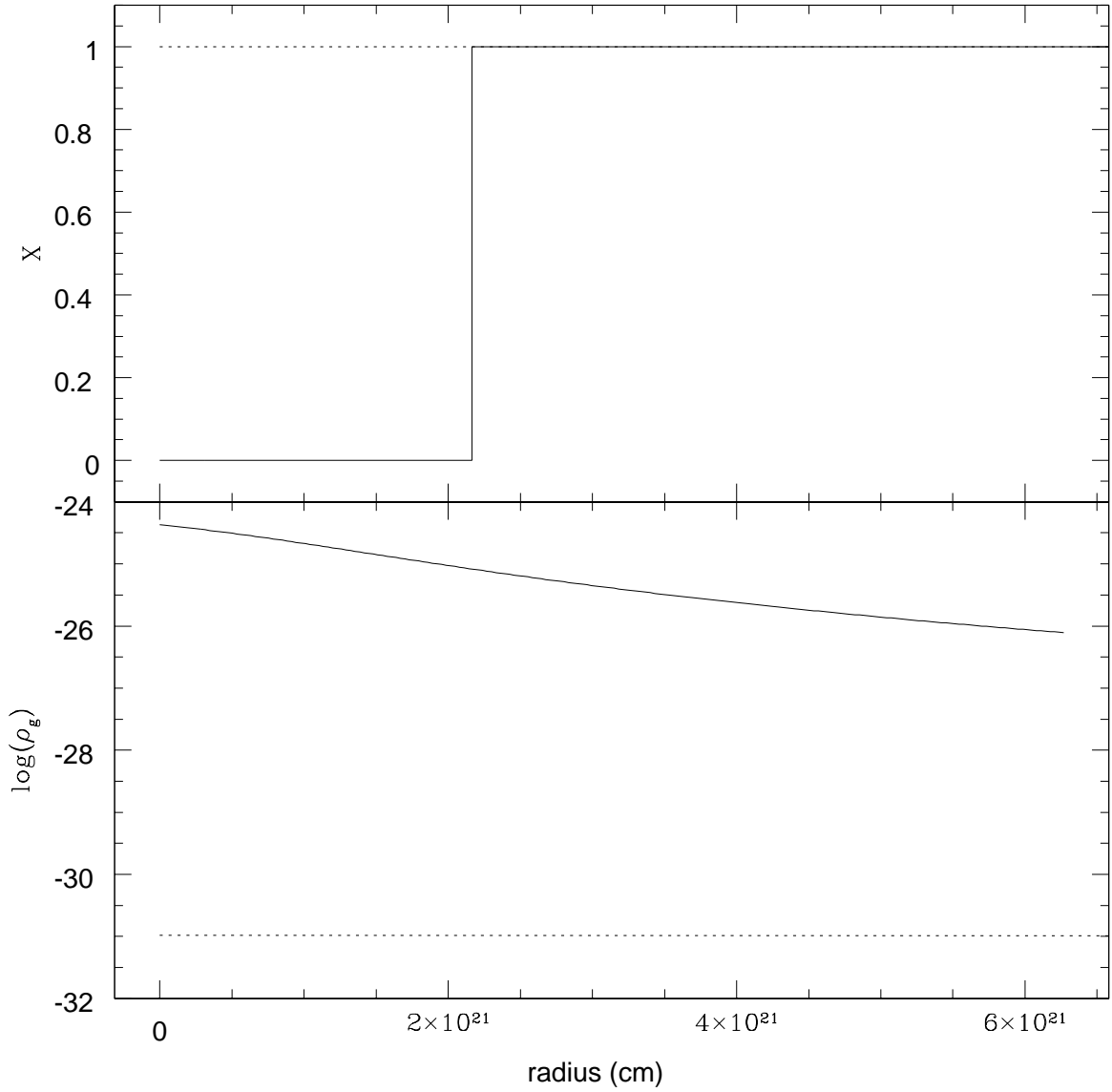}
\caption{Ionization fraction and density profile for the initial
(solid lines) and final (dotted lines) states of model A2.}
\label{fig:a2xd}
\end{figure}

For models A2 and A3, the outer regions of the galaxy (beyond 0.6 and
1~kpc respectively), quickly become photoionized while the gas that is
initially located near the central regions of the galaxy collapses to
form stars with $f_\star = 3.6 \times 10^{-2}$ and $5.2 \times
10^{-2}$.  The total stellar content is
\begin{equation}
M_\star = f_\star b M_{DM},
\end{equation}
so that the total mass of the stars formed in these two models are
$4.6 \times 10^5 M_{\sun}$ and $1.3 \times 10^6 M_{\sun}$, comparable
to stellar content of the Ursa Minor and Leo I dSph respectively.
The results of these models
indicate that the hydrogen gas has sufficiently large column density
to become self-shielded against the external UV flux if the ionization
fraction is low.  But, in both models, the onset of star formation
quickly leads to photoionization from stellar UV emission, and an
outward expansion of the gas.  In the shallow potential of
low-velocity dwarf galaxies, a large fraction of the ionized gas
escapes from the system.  Eventually, the residual gas evolves toward
new equilibria, with a density that is too low for effective
recombination.  Consequently, an inflated, hot, and highly ionized
state is attained, in which star formation is quenched
(Figure~\ref{fig:a2xd}).

The comparison between models A1, A2, and A3 indicates that
unperturbed dwarf galaxies with very low ordinary matter contents are
likely to remain fully ionized.  Nevertheless, a small population of
stars may be formed within these galaxies if a critical amount of gas
can become neutral.  The conversion efficiency is a function of the
baryonic mass.  Hydrogen may become neutral either due to initial
conditions, by accretion of additional gas from the intergalactic
medium, or it may be induced by cloud-cloud collisions (see below).
Although mass loss can reduce $M_\star/M_{DM}$ to well below both its
initial value and the cosmological $\Omega_b/\Omega_d$, the amount of
stars formed in these systems provide a lower limit upon its initial
value, $b$.

\subsubsection{Very Low Mass Dwarf Galaxies}

Although the potential in models A1, A2, and A3 is relatively small,
most dwarf spheroidal galaxies have even shallower potentials and
smaller velocity dispersions.  In order to assess the efficiency of
initial star formation in such systems, we consider models B1, B2, and
B3 , in which $r_0$ is set to be 0.3, such that $v_0 = 8.5$~km
s$^{-1}$, $M_0 = 4.3 \times 10^6 M_{\sun}$, and $M_{DM} = 2.5 \times
10^7 M_{\sun}$.  The value of $b$ for these three models are 0.1, 0.2,
and 0.5, so that the total baryonic mass in model B3 equals that in
model A2.  The external UV flux all three models is $J_0 = 4. \times
10^{-23} \rm{erg \ s^{-1} \ cm^{-2} \ ster^{-1} \ Hz^{-1}}$.

In comparison with model A2, model B3 is more compact with a higher
$\rho_0$ and more rapid decline in the density distribution.  Although
the actual baryonic mass contained in the initial Str\"omgren radius
in model B3 is smaller than that in model A2, the gravitational
potential for $r_0=0.3$ is shallower than that for $r_0=0.6$.  In both
models, the gas was not in a hydrostatic equilibrium initially.
Because the gas collapses faster in the more massive model A2 than in
model B3, a slightly larger fraction of the gas was turned into stars
during the initial formation episode.

\begin{figure}
\plotone{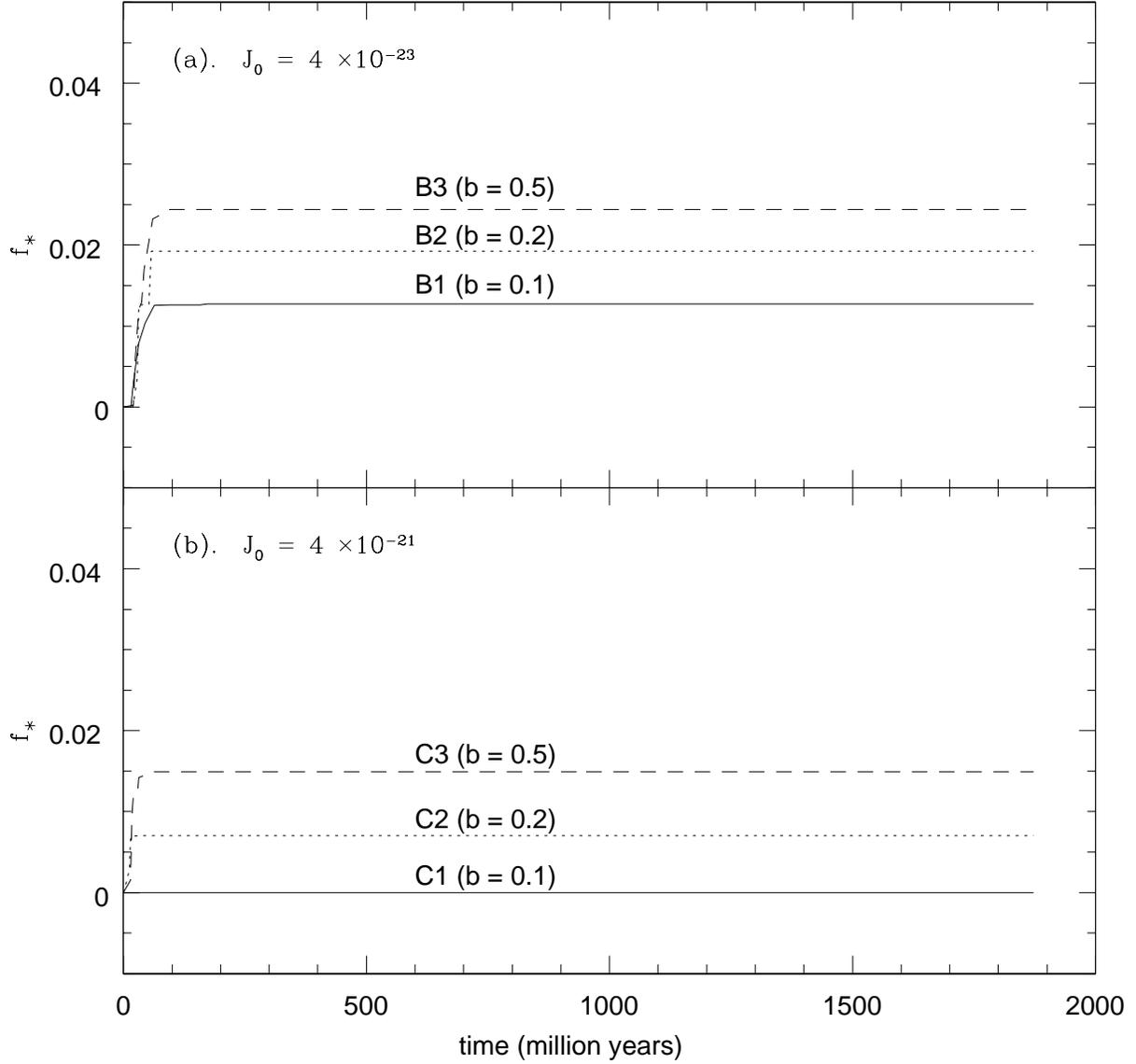}
\caption{Star formation history for models with $r_0 = 0.3$~kpc: (a). $J_0
= 4 \times 10^{-23} $, and $b = 0.001, 0.1$ and $0.2$ for
model B1, B2 and B3 respectively; (b). $J_0 = 4 \times 10^{-21} $, and
$b = 0.001, 0.1$ and $0.2$ for model C1, C2 and C3 respectively.}
\label{fig:r003}
\end{figure}

The subsequent evolution of the gas and stars in models B1, B2, and B3
are similar to those in models A2 and A3.  After an initial episode of
star formation, the residual gas becomes fully ionized, a substantial
amount of mass is lost, and star formation is quenched.  The asymptotic
values of $f_\star$ for these models are plotted in
Figure~\ref{fig:r003}a.  The trend of $f_\star$ increasing with $b$
generally holds.  The magnitude of $M_\star$ is $8 \times 10^3
M_{\sun}$ and $2\times 10^4 M_{\sun}$ for models B2 and B3
respectively.  Thus, even with $b$ substantially larger than the
cosmological $\Omega_b/\Omega_d$, the mass of stars formed is
substantially below the stellar contents in dSphs.

\subsubsection{Star Formation at Early Epochs}

In the above models, we adopt the background UV flux for low redshift.
The old stellar population in dSphs was, however, formed at high
redshifts, when the background UV flux was considerably more intense.
In models C1, C2, and C3 (Figure~\ref{fig:r003}b), we adopt $J_0 =
4\times 10^{-21} {\rm ergs \ s^{-1} \ cm^{-2} \ ster^{-1} \ Hz^{-1}}$,
with $r_0 = 0.3$, and, respectively, $b=0.1, 0.2,$ and $0.5$.  All
other model parameters are similar to those in the B series of models.
Although $J_0$ in models C1, C2 and C3 is $10^2$ larger than that in
the previous models, the resulting $f_\star$ and hence $M_\star$ for
models B3 and C3 are comparable to each other.  The density at the
Str\"omgren surface is proportional to $J_0^{1/2}$, and so the
C-series of models are photoionized closer to the galaxy center.  But
the initial episode of star formation in both model B3 and C3 is
quenched primarily by the feedback, rather than by photoionization
from external UV radiation.  The difference in $f_\star$ is
substantial, however, between the most tenuous models, B1 and C1.  The
value of $J_0$ is sufficient to completely ionize the gas in
model C1 prior to the onset of collapse.  Consequently, the star
formation efficiency is essentially zero for model C1.

\subsection{Effect of Gravitational Potential}

\subsubsection{Star Formation at Recent Epochs}

The models considered above have relatively shallow gravitational
potentials, such that the residual gas can escape after it is ionized.
We now consider a series of models with deeper gravitational
potentials.  In models D1, D2, D3, D4, and D5, we adopt, respectively,
$r_0=0.3$, $0.6$, $2$, $4$, and $6$.  We set $b=0.1$ for all five
models.  In each, we also use $J_0 = 4 \times 10^{-23} \rm{ergs \
s^{-1} \ cm^{-2} \ ster^{-1} \ Hz^{-1}}$, such that models D1 and D2
are identical, respectively, to models B1, and A2.  For models D3, D4,
and D5, respectively, $M_0= 3.6\times 10^8 M_{\sun}$, $1.8 \times 10^9
M_{\sun}$, and $4.7 \times 10^9 M_{\sun}$; $M_{DM} = 2.1 \times 10^9
M_{\sun}$, $10^{10} M_{\sun}$, and $2.7 \times 10^{10} M_{\sun}$; and
$v_0 = 28$, $45,$ and $58 \rm{\ km \ s^{-1}}$.  The potentials of these
models are deeper than those of the dSphs but comparable to those of
typical dwarf elliptical and dwarf irregular galaxies.

\begin{figure}
\plotone{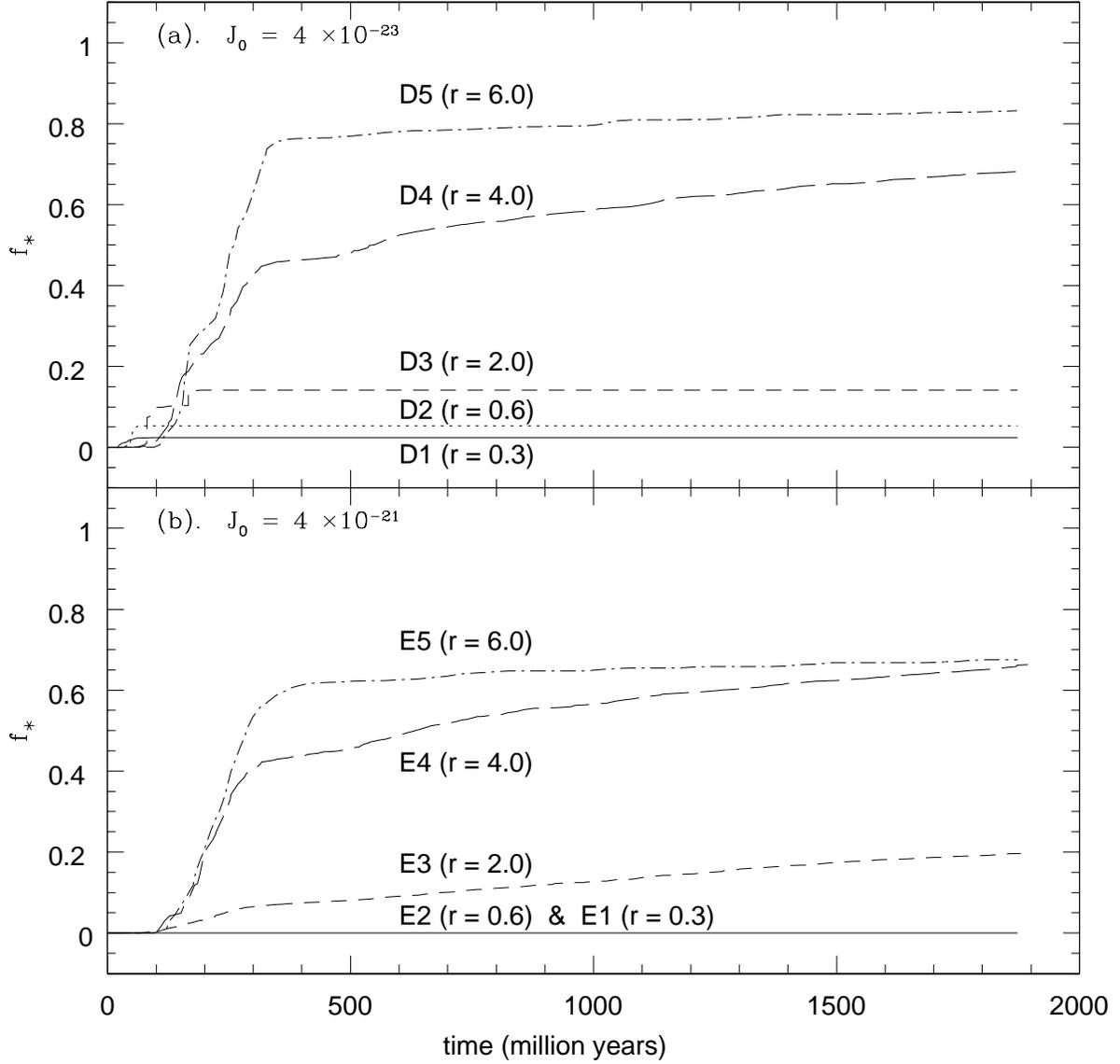}
\caption{Star formation history for models with $b = 0.1$: (a). $J_0
= 4 \times 10^{-23} $, and $r_0 = 0.3, 0.6, 2.0, 4.0$ and $6.0$ for
model D1, D2, D3, D4 and D5 respectively; (b). $J_0
= 4 \times 10^{-21} $, and $r_0 = 0.3, 0.6, 2.0, 4.0$ and $6.0$ for
model E1, E2, E3, E4 and E5 respectively.}
\label{fig:b01}
\end{figure}

In Figure~\ref{fig:b01}a, we plot the evolution of $f_\star$ for
various models.  These results clearly indicate that the residual gas
is retained by the gravitational potential of the dwarf galaxies in
models D4 and D5.  Star formation in each of these cases is
self-regulated.  In model D4, the mass of the stellar component
reaches $M_\star \sim 6.4 \times 10^8 M_{\sun}$ with a comparable
amount of residual gas bound to the galaxy after 1~Gy.  Thereafter,
star formation continues at a self-regulated rate of $\sim 0.2{\rm \
M}_{\sun}$~yr$^{-1}$.  For the more massive model D5, more than half
of its original gas is converted into stars relatively early.  After
$300$ Myr, the mass of the stars becomes $2 \times 10^9{\rm \
M}_{\sun}$ with $\sim 5\times 10^8{\rm \ M}_{\sun}$ of residual gas.
The residual gas is much more extended than originally, which slows
down the star formation rate.

\subsubsection{Star Formation at Early Epochs}

The external UV flux in models D1-5 corresponds to the present-day
value.  In order to consider the conditions under which stars as old
as 10-12~Gyr formed, we consider models E1-5 (Figure~\ref{fig:b01}b).
These models are identical to models D1-5, except that $J_0$ is
increased by two orders of magnitude ({\it i.e.} $J_0 = 4 \times
10^{-21}\ {\rm erg}\ {\rm s}^{-1} {\rm cm}^{-2}{\rm ster}^{-1}{\rm
Hz}^{-1}$).  With this enhanced UV flux, the gas in model E2 is ionized
immediately, and no star formation occurs.  By contrast,
the lower UV flux present in model D1 allows for the initial formation
of a few stars.

\begin{figure}
\plotone{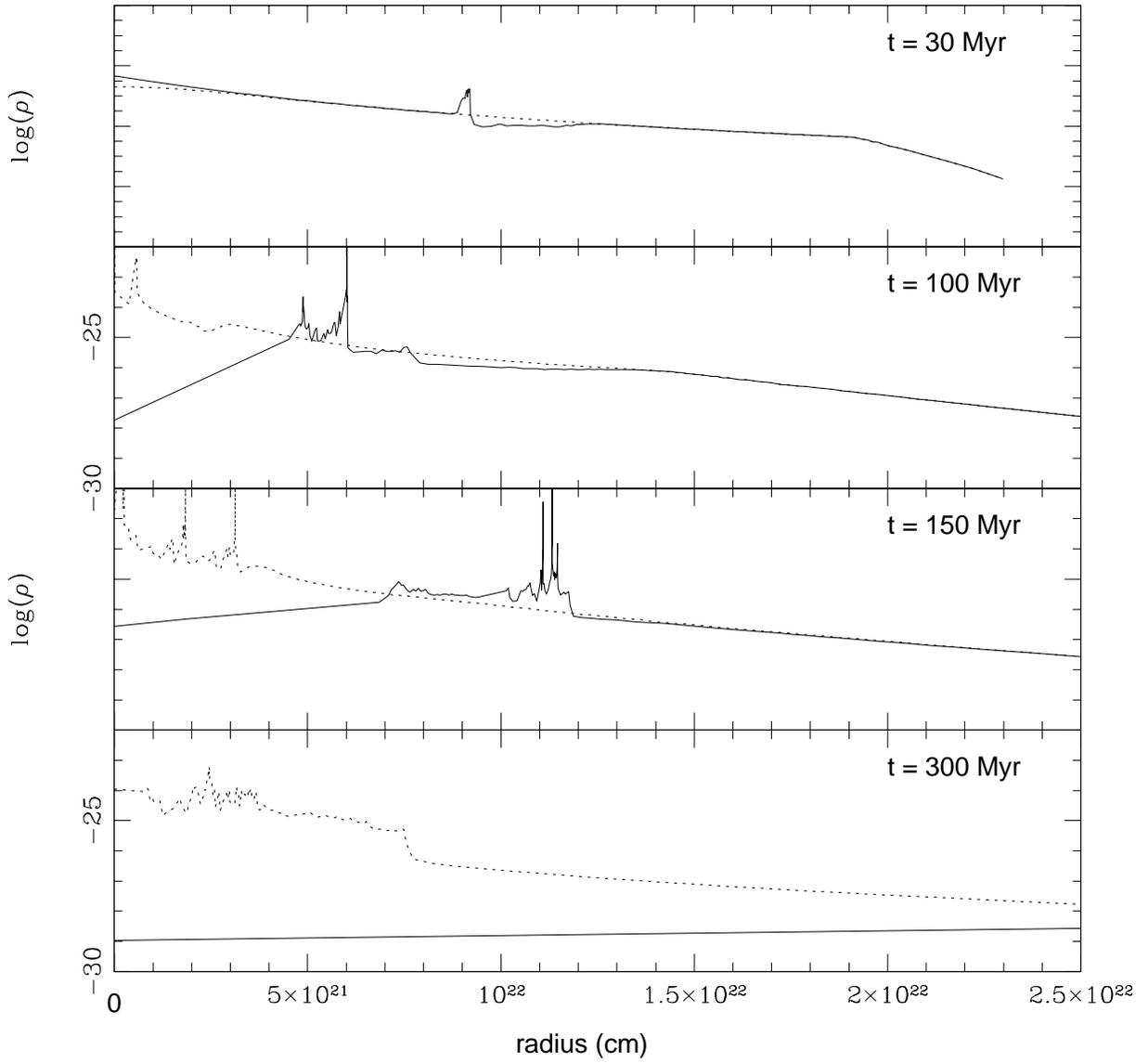}
\caption{Gas density profile snapshots for model D3 (solid lines) and E3
(dotted lines) at $t = 30, 100, 150$ and $300$ Myr.}
\label{fig:snap}
\end{figure}

The intense UV radiation increases the density needed for ionization
equilibrium, and so the radius of the Str\"omgren sphere is reduced.
This effect is most evident in the marginal models D3 and E3.
Photoionization due to the background is weak in the low $J_0$
model~D3.  In that model, the residual gas undergoes rapid collapse,
leading to high early levels of star formation and feedback, resulting
in two early, strong `bursts' of star formation.  The momentum
injected to the gas by the star formation is sufficient to drive the
gas outwards, such that it becomes too tenuous to recombine, and
subsequent star formation is quenched.  In contrast, the outer regions
of model~E3 are ionized.  But the gravitational potential is
sufficiently deep to retain the ionized gas.  Star formation is
therefore able to proceed, and does so at a nearly constant rate,
regulated by the supernova injection of momentum, and reduced relative
the initial rate of model~D3 due to the higher level of ionization.
Somewhat counterintuitively, therefore, model~E3 by late times is able
to convert a higher fraction of its gas into stars than is model~D3.
In Figure~\ref{fig:snap}, we show the density profile snapshots of
model D3 (solid lines) and E3 (dotted lines) at
$t = 30, 100, 150$ and $300$ Myr.

Evolution of the star formation rate for the more massive models E4
and E5 are nearly indistinguishable from those of D4 and D5.  This
implies that the rich supply of gas in both cases limits the effects
of the external UV radiation to the gas in the tenuous outer halo,
well outside of the central, star-forming regions, which have much
shorter recombination and dynamical timescales.

\begin{figure}
\plotone{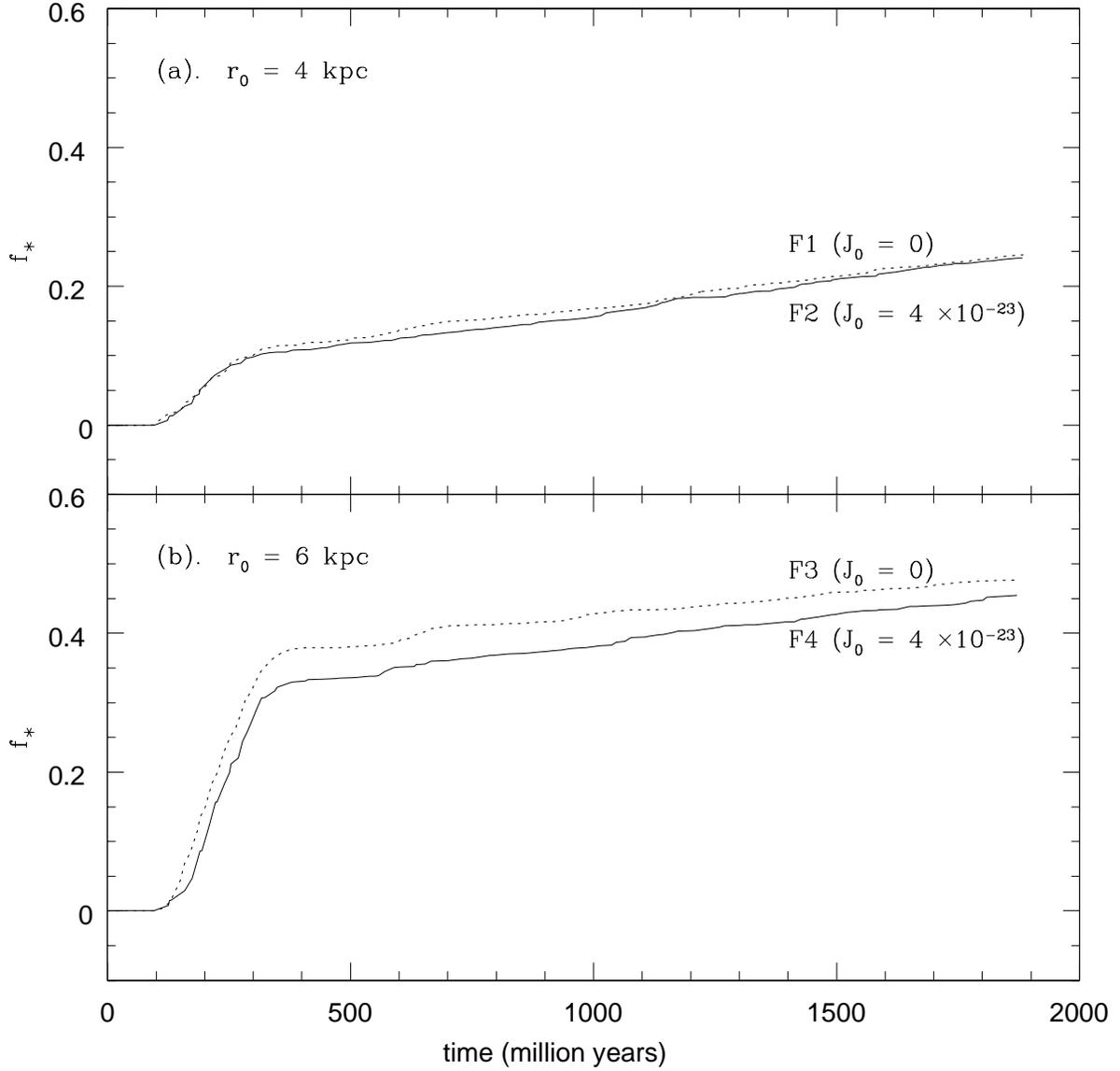}
\caption{Star formation history for models with $b = 0.01$.
(a). $r_0 = 4.0$, and $J_0 = 0, \  4 \times 10^{-23} $ for
model F1, F2 respectively;
(b). $r_0 = 6.0$, and $J_0 = 0, \ 4 \times 10^{-23} $ for
model F3, F4 respectively.}
\label{fig:b001}
\end{figure}

In order to verify this tendency, we examine models F1-F4, in all four
of which $b =0.01$.  In models F1 and F2, $r_0=4$~kpc, whereas it is
6~kpc for models F3 and F4.  The external $J_0 = 4 \times 10^{-23}\
{\rm erg}\ {\rm s}^{-1}{\rm cm}^{-2}{\rm ster}^{-1}{\rm Hz}^{-1}$ for
models F1 and F3, whereas $J_0=0$ for models F2 and F4.  The evolution
of $f_\star$ for all four models is shown in Figure~\ref{fig:b001}.
The comparison between models F1 and F2 (as well as that between F3
and D4) indicates that the external flux does not significantly modify
the star formation rate.  The nearly steady star formation rate is
self-regulated.

\subsection{Star Formation Efficiency}

\begin{figure}
\plotone{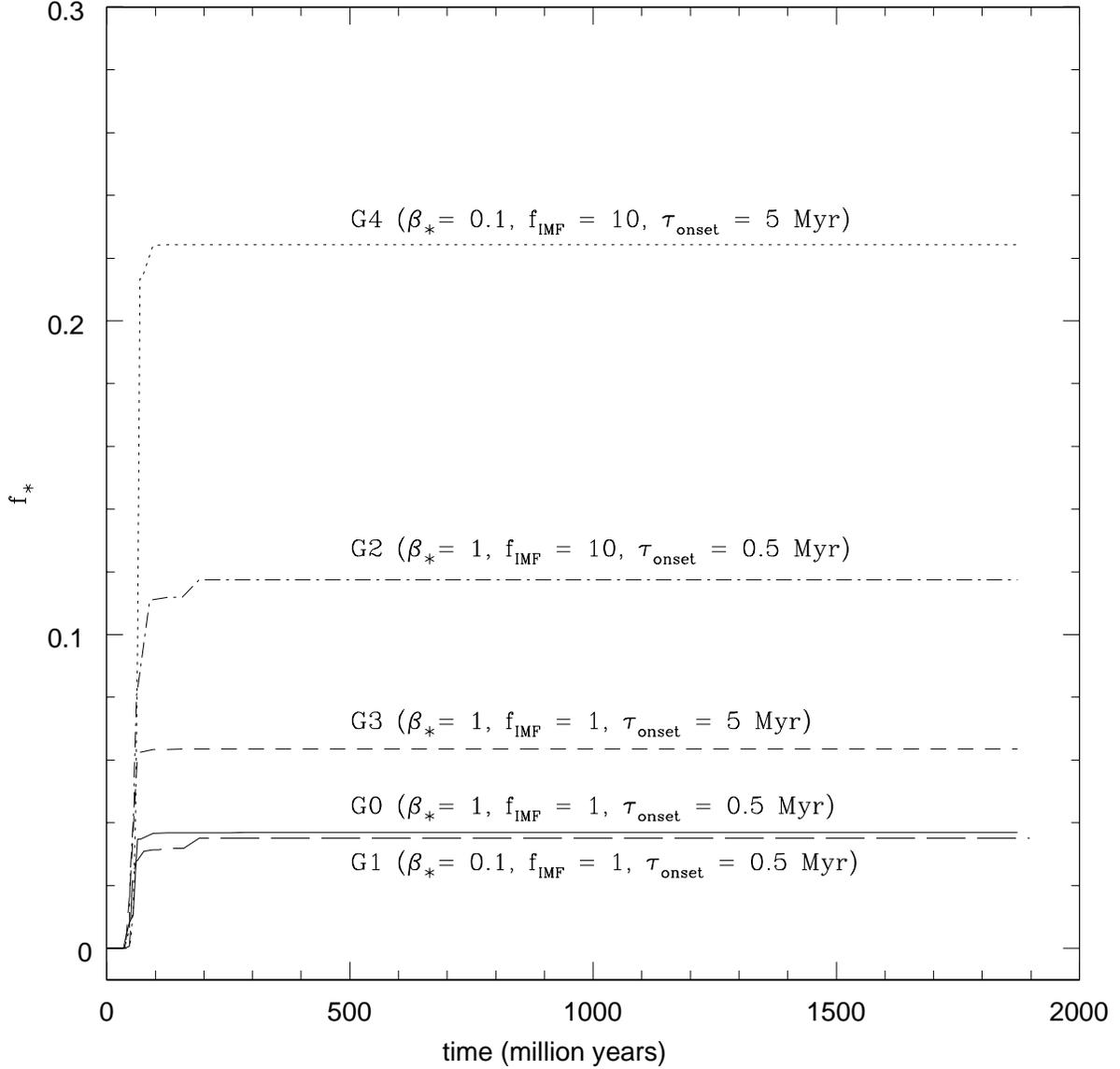}
\caption{Star formation efficiency.  All models with $r_0 = 0.6$~kpc, $J_0
= 4 \times 10^{-23}$, and $b = 0.2$, but with different parameters for
star formation efficiency. Model G0 (base model): $\beta_{max} = 1$, $f_{IMF}
= 1$, $\tau_{onset} = 0.5$~Myr; Model G1: $\beta_{max} = 0.1$, $f_{IMF}
= 1$, $\tau_{onset} = 0.5$~Myr; Model G2: $\beta_{max} = 1$, $f_{IMF}
= 10$, $\tau_{onset} = 0.5$~Myr; Model G3: $\beta_{max} = 1$, $f_{IMF}
= 1$, $\tau_{onset} = 5$~Myr; Model G4: $\beta_{max} = 0.1$, $f_{IMF}
= 10$, $\tau_{onset} = 5$~Myr}
\label{fig:param}
\end{figure}

The above results indicate that, in systems with $r_0 < 2$ kpc, the
residual gas expands towards a new equilibrium after an initial epoch
of star formation.  Consequently, mass is lost from the galaxy and the
residual gas density becomes sufficiently depleted that the external
background UV flux can severely suppress the star formation rate.  The
magnitude of $f_\star$ may be enhanced if the feedback effect is weaker
than in the standard models.  Such might be the case if $\beta_{max}$
is relatively small (corresponding to an inefficient conversion of gas
into stars), or if $f_{IMF}$ is relatively large (corresponding to a
suppressed production rate of the massive supernovae progenitor
stars).  In models G1-G4, we consider some variations of model A3
(which is also referred to as G0 for comparison purposes).  In each,
$J_0 = 4 \times 10^{-23}\ {\rm erg}\ {\rm s}^{-1}{\rm cm}^{-2} {\rm
ster}^{-1}{\rm Hz}^{-1}$, $b=0.2$, and $r_0 =0.6$~kpc.  In model~G1,
we set $\beta_{max} = 0.1$ and $f_{IMF} = 1$.  In models~G2 and G3, we
set $\beta_{max} = 1$ and $f_{IMF} =10$, while in model~G4, $\beta_{max}
=0.1$ and $f_{IMF} = 10$.  In models~G0-2, $\tau_{onset}=0.5$~Myr, but
in models~G3 and G4, $\tau_{onset} = 5$~Myr (see
Fig.~\ref{fig:param}).  The magnitude of $f_\star$ for model G3 is
almost an order of magnitude larger than that of model~G0 (A3).  In
comparing between models~G1 and G2, we see that $f_\star$ depends more
sensitively on the value of $\beta_{max}$ than upon $f_{IMF}$ and
$\tau_{onset}$.

\subsection{Self-regulated, Episodic Star Formation}

Star formation in the more massive galaxies, with larger $r_0$,
discussed above is
self-regulated by the feedback process.  In the models
presented so far, self-regulated star formation leads to
quasi-equilibria, such that the star formation rate is essentially
constant.  For models D4, D5, E4, and E5 there are small variations in
the star formation rate during the early evolution of the galaxy.  In
these models, the star formation rate is relatively steady, because the
local values of $\tau_{ff} + \tau_{onset}$ are much shorter than the
average dynamical timescale of the star forming region, which itself
is regulated by the feedback process.  In contrast, for model D3, with
a more compact structure, the time delay can lead to episodic
variations in the star formation rate.  However, the supply of
baryonic mass is limited, and after one or two episodes of star
formation the density is reduced sufficiently that the residual gas
remains ionized and star formation is quenched.

Some dwarf galaxies show evidence for episodic star formation
histories.  In order for episodic star formation to take place,
three conditions must be satisfied: 1) the galaxy must have sufficient
baryonic mass to become self-shielded against the background UV
radiation, 2) the collapse and feedback switch-on delay times must be
longer than the global dynamical timescale, in order that a
substantial amount of baryonic matter may accumulate prior to the
onset of each episode of star formation, and 3) the star formation
efficiency must be modest, so that an adequate amount of residual gas
may be preserved for the subsequent episodes of star formation.
Models H1-H5, have the same values of $r_0 (=4$~kpc$)$, and $J_0$ as
model D4 (which we aso refer to as the fiducial model~H0).  In models H1
and H2, we set $\tau_{onset} = 5$ My and 50 My respectively.  In both
cases, the delay in the switch on of the feedback effects allows more
material to accumulate into the central regions of the galaxy.  The
delay enhances the eventual rate of star formation leading to ferocious
feedback, driving all the residual gas out of the galaxies.

\begin{figure}
\plotone{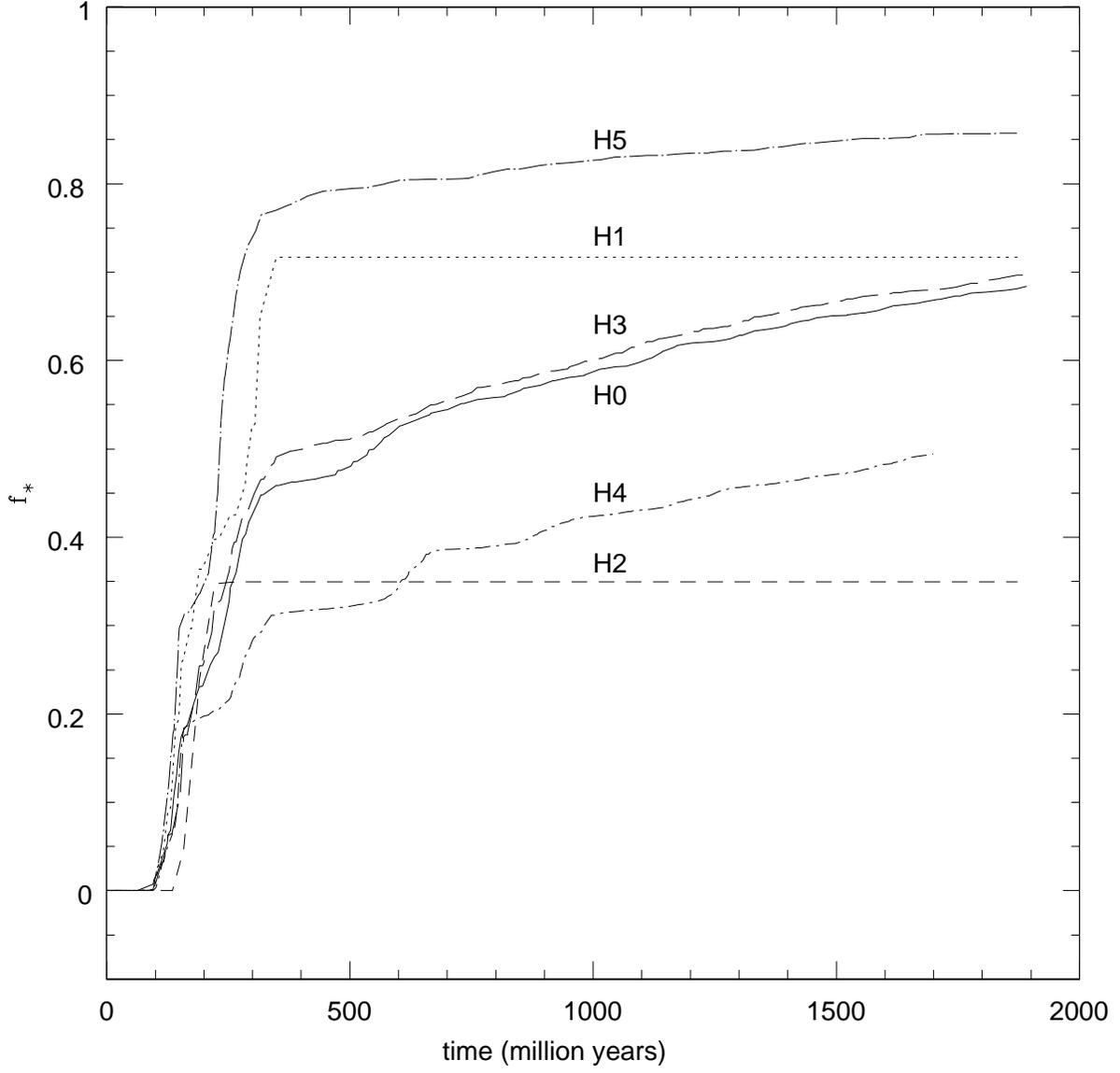}
\caption{Star formation efficiency.  All models with $r_0 = 4.0$~kpc, $J_0
= 4 \times 10^{-23}$, and $b = 0.1$, but with different parameters for
star formation efficiency. Model H0 (base model): $\beta_{max} = 1$, $f_{IMF}
= 1$, $\tau_{onset} = 0.5$~Myr; Model H1: $\beta_{max} = 1$, $f_{IMF}
= 1$, $\tau_{onset} = 5$~Myr; Model H2: $\beta_{max} = 1$, $f_{IMF}
= 1$, $\tau_{onset} = 50$~Myr; Model H3: $\beta_{max} = 0.1$, $f_{IMF}
= 1$, $\tau_{onset} = 5$~Myr; Model H4: $\beta_{max} = 0.01$, $f_{IMF}
= 1$, $\tau_{onset} = 5$~Myr; Model H5: $\beta_{max} = 0.1$, $f_{IMF}
= 10$, $\tau_{onset} = 5$~Myr}
\label{fig:r4par}
\end{figure}

In models H3 and H4, we set, respectively, $\beta_{max} = 0.1$ and
$10^{-2}$, motivated to satisfy requirement 3) above for episodic star
formation.  In Figure~\ref{fig:r4par}, we see that these models indeed
lead to episodic variations in the star formation rate.  Finally, in
model H5, we also include the effect of a bottom-heavy initial mass
function by setting $f_{IMF} = 10$.  In this case, a large fraction of
the ordinary matter is retained by the galaxies, and converted into
stars.

\subsection{Effect of External Perturbations}

One difficulty associated with the results presented above
is the low gas-to-star conversion efficiency for the lowest
mass dwarf galaxies.  In addition to reducing the severity of the
feedback effect, we now explore the possibility of compressions, such
as would result from either tidal perturbations or collisions with
other galaxies.  The models which we consider are shown in
Table~\ref{tab:galaxymodels2}.  In that table are listed, for each model,
the values of $r_0$, $b$, $J_0(z)/J_0(0)$, $H_p/H_0$, $f_{IMF}$,
$\tau_{onset}$, and $f_\star$.  The value of $H_p$ indicates the strength
of a linear compression imposed upon the models, as described in
Equation~\ref{eqn:compeqn}, and is normalized in the table to
$H_0 = 1$~km~s$^{-1}$~kpc$^{-1}$.

Models I1-I4 have values of $J_0$, $r_0$, and $b$ identical to those
of model~A3.  In models I1-I4, respectively, $H_p = 3, 15, 30,$ and
150~km~s$^{-1}$~kpc$^{-1}$ ($h_p = 0.49, 2.4, 4.9$, and $24$).
The compressional timescale across the galaxies, $r_{halo}/H_p = $ 1,
0.2, 0.1, and 0.02~Gyr, respectively, for models~I1-4. For the region
which contains most of the gas in models I3 and I4, the compressional
speed is supersonic and the density jump at the shock fronts induces
large increases in the local star formation rate and efficiency.
Consequently, $f_\star$ is substantially enhanced.

\begin{figure}
\plotone{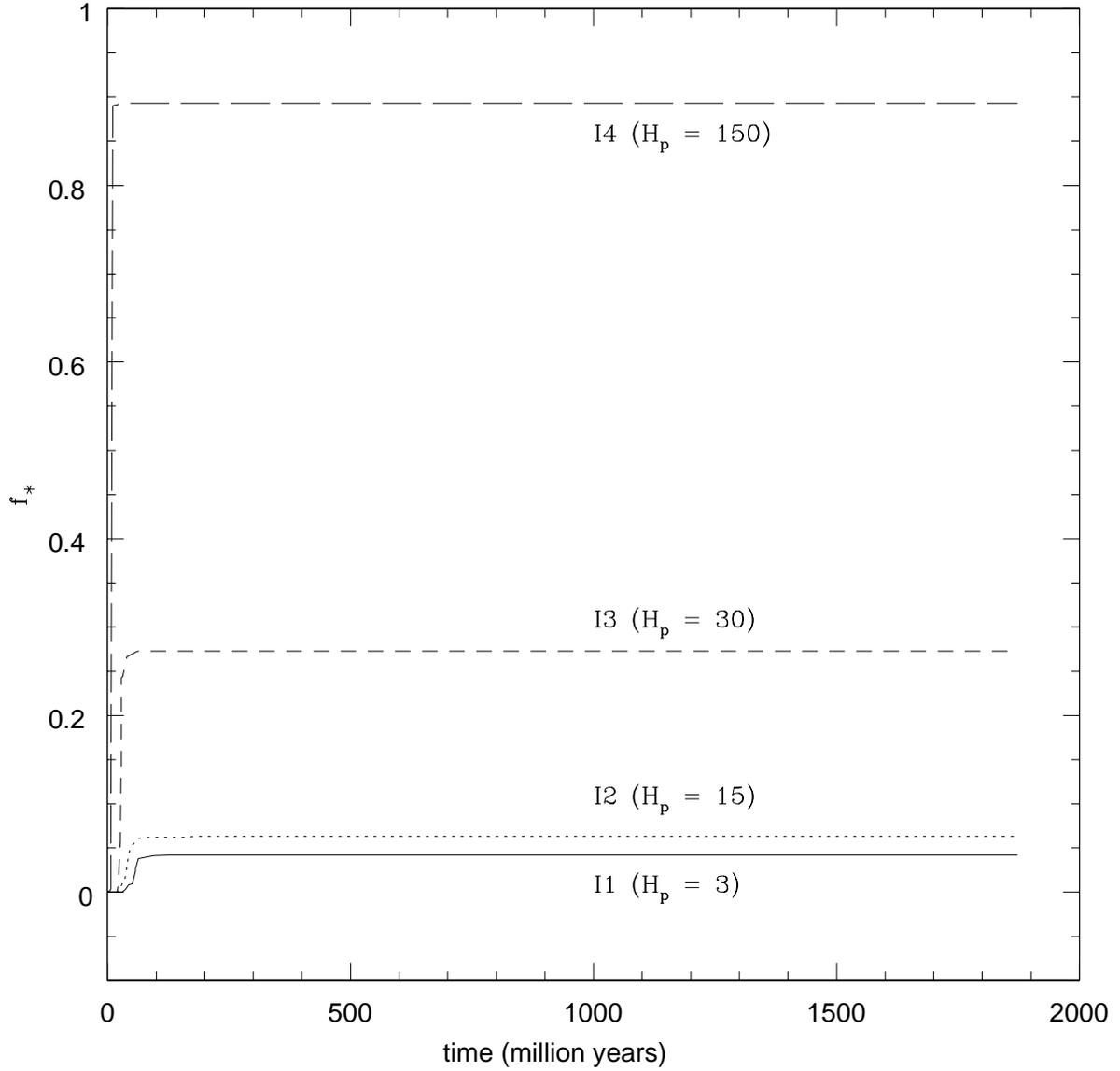}
\caption{Star formation history for models with external
perturbation. All models with $r_0 = 0.6$~kpc, $J_0
= 4 \times 10^{-23}$, and $b = 0.2$, but with different external
perturbation strength: $H_p = 3, 15, 30,$ and
150~km~s$^{-1}$~kpc$^{-1}$ for models I1-I4 respectively.}
\label{fig:hubble}
\end{figure}

The results are shown in Figure~\ref{fig:hubble}, where it can clearly
be seen that $f_\star$ is an increasing function of $H_p$.  The
enhancement is associated with the increased densities, and resulting
enhanced recombination rates, which provide a better shield for the
residual gas in the galaxy.  We note that the compression
is excessive for model~I4.  Such a large impact is more appropriate for
cloud-cloud collisions, rather than tidal compression.  These results
indicate a significant fraction of the baryons in dwarf galaxies can
be converted into stars if the galaxies are subjected to supersonic
compressions.

\begin{figure}
\plotone{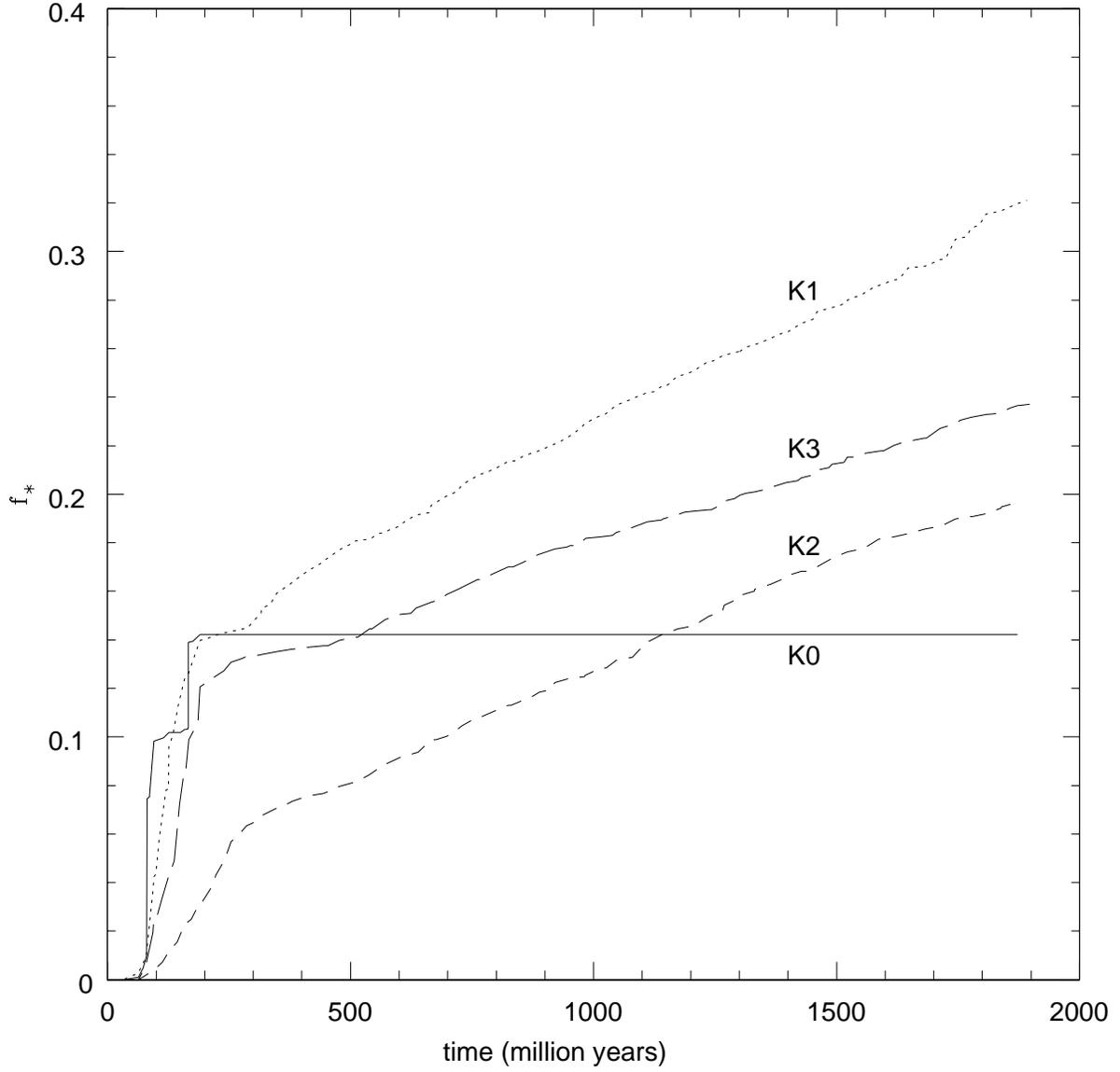}
\caption{Effect of external perturbation. All models with
$r_0 = 2$~kpc, and $b = 0.1$. Same $J_0
= 4 \times 10^{-23}$, but different $H_p = 0$ and $3$ for models K0 and K1
respectively; Same $J_0
= 4 \times 10^{-21}$, but different $H_p = 0$ and $3$ for models K2 and K3
respectively.}
\label{fig:hbblr2}
\end{figure}

Finally, we consider the effect of impact on relatively massive dwarf
galaxies similar to models D3 and E3 (which is refereed to as models K0
and K2 here). We compare the results of these models without external
perturbation to those of models K1 and K3 with a modest external
perturbation ($H_p = 3$~km~s$^{-1}$~kpc$^{-1}$).
Both K0 and K1 have an external
UV flux of $J_0 = 4 \times 10^{-23}\ {\rm erg}\ {\rm s}^{-1}{\rm cm}^{-2} {\rm
ster}^{-1}{\rm Hz}^{-1}$; while both K2 and K3
$J_0 = 4 \times 10^{-21}\ {\rm erg}\ {\rm s}^{-1}{\rm cm}^{-2} {\rm
ster}^{-1}{\rm Hz}^{-1}$. The results are
shown in Figure~\ref{fig:hbblr2}. For models K1 and K3, the
compression speed is
slightly slower than the sound speed of the ionized gas in the outer
regions of the galaxies.  Nevertheless, it can confine the thermally
expanding ionized shell.  Thus, the compressional speed not only leads
to a density enhancement, but also retention of the gas, enabling the gas
in model K1 to undergo protracted and self-regulated star formation.
We also note that even with such a modest compressional speed, gas in
these dwarf galaxies can recombine efficiently even in early
cosmological epoches when the intensity of the UV flux is relatively
high.  The comparison between models K0 and K1 indicates that this
compression can double the star formation efficiency.

\section{Summary and Discussion}

In this paper, we consider a series of one-dimensional numerical models
to examine the efficiency of star formation in dwarf galaxies under the
influence of local cooling, heating by UV photons from both the background
radiation and hot stars, and the winds and supernovae associated with massive
stars.  Radiative transfer is handled using a simple ``Str\"omgren shell''
approximation.  Completely ionized gas is assumed to have temperature of
15,000~K, while heating and cooling are followed dynamically in gas that is
not completely ionized.  The effects of supernovae and winds from massive
stars are included in the momentum equation of the gas.  Stars can form in
gas that has cooled sufficiently for the Jeans mass to be within the range of
stellar masses, which is not tidally limited, and for which the free-fall
timescale is short compared to the pre-main sequence evolution time of
massive stars, $\tau_{onset}$.  The rate of conversion of gas into stars is
a fraction of the gas free-fall timescale.  Feedback from the massive stars
which form is delayed by the free-fall time, plus $\tau_{onset}$.

The most important parameters explored in the models are: 1) the
depth of the galactic potential, 2) the baryonic fraction, 3) the
background UV flux, 4) $\tau_{onset}$, 5) the IMF, and
6) compressional perturbation (see Tables 1 and 2).

\subsection{The Initial Baryonic-to-Dark Matter Ratio}
We initially confirm the results of Kepner et al (1997), that for a
low mass system ($M_{DM}\lesssim10^8{\rm\ M}_\odot$) with baryonic
fraction $b = 0.001$, even today's background UV flux is adequate to
totally ionize the gas such that no stars can form.  With increasing
baryonic fraction, however, these systems eventually become
self-shielded from the background UV, and the gas is able to cool and
collapse, leading to star formation.  For $b=0.2$ (comparable to the
ratio of $\Omega_b/\Omega_d$ today), a low mass system with $r_0 =
0.3$~kpc is able to convert $\sim 2\%$ gas ($\sim 10^5\ M_{\sun}$)
into stars in about 2 billion years.  Feedback from star formation
leads to rapid expansion of the gas within the shallow galactic
potential.  The lower recombination efficiency of the expanded gas
allows it to be completely ionized by the external UV radiation, and
star formation is terminated after the initial, weak burst.  Modest
retention of gas is possible, if the IMF of the stars formed in these
system is biased towards the low-mass stars.

In the presence of much stronger background UV radiation, appropriate
to early epochs, star formation may still occur in small galaxies
having $b\gtrsim0.2$.  In such systems, self-shielding is sufficient
to prevent ionization initially within the cores, even with the higher
UV background.  Star formation then proceeds, governed solely by
internal time scales and feedback, while the external radiation acts
only at late times, to prevent further star formation.  The
requirement for a relatively large concentration of ordinary matter in
these shallow potential is in contrast to the low-$b$ estimate made by
Kepner {\it et al} (1997) based on the assumption of Bondi accretion
into dark matter potentials.  Although the accretion process may be
more effective when the pristine gas is cold before the epoch of
reionization, all present-day dwarf galaxies must have acquired their
ordinary matter content prior to the formation of massive stars in
other nearby dwarf galaxies.  The relatively short dynamical time
scales within the dwarf galaxies suggests that such a precise timing
may be difficult to achieve.  Alternatively, prior to
the epoch of dwarf galaxy formation, density fluctuation of the
baryonic matter may be well correlated and coupled to that of the dark
matter.

\subsection{Mass-to-Light Ratio}

Star formation is much more efficient in higher mass systems, due both
to increased self-shielding and to the enhanced ability of the
galactic potential to retain ionized gas.  A ten-fold increase in the
dark matter mass, from $10^{7-8}$ typical of dSphs, to $10^{8-9}$
typical of dwarf Ellipticals (dEs) or dwarf Irregulars (dIrrs), may
lead to an order of magnitude increase in the gas-to-star conversion
efficiency.  In the most massive systems, the star formation rate is
self-regulated by internal feedback, independent of the external UV
flux.  For marginal systems, with low external flux, star formation
occurs in a rapid initial burst, followed by expansion and ionization
by the external flux, similar to the low-mass systems.  The presence
of a higher external UV flux, however, lessens the initial star
formation rate, reducing the subsequent feedback, and allowing star
formation to proceed at a slow pace for an extended period.

\begin{figure}
\plotone{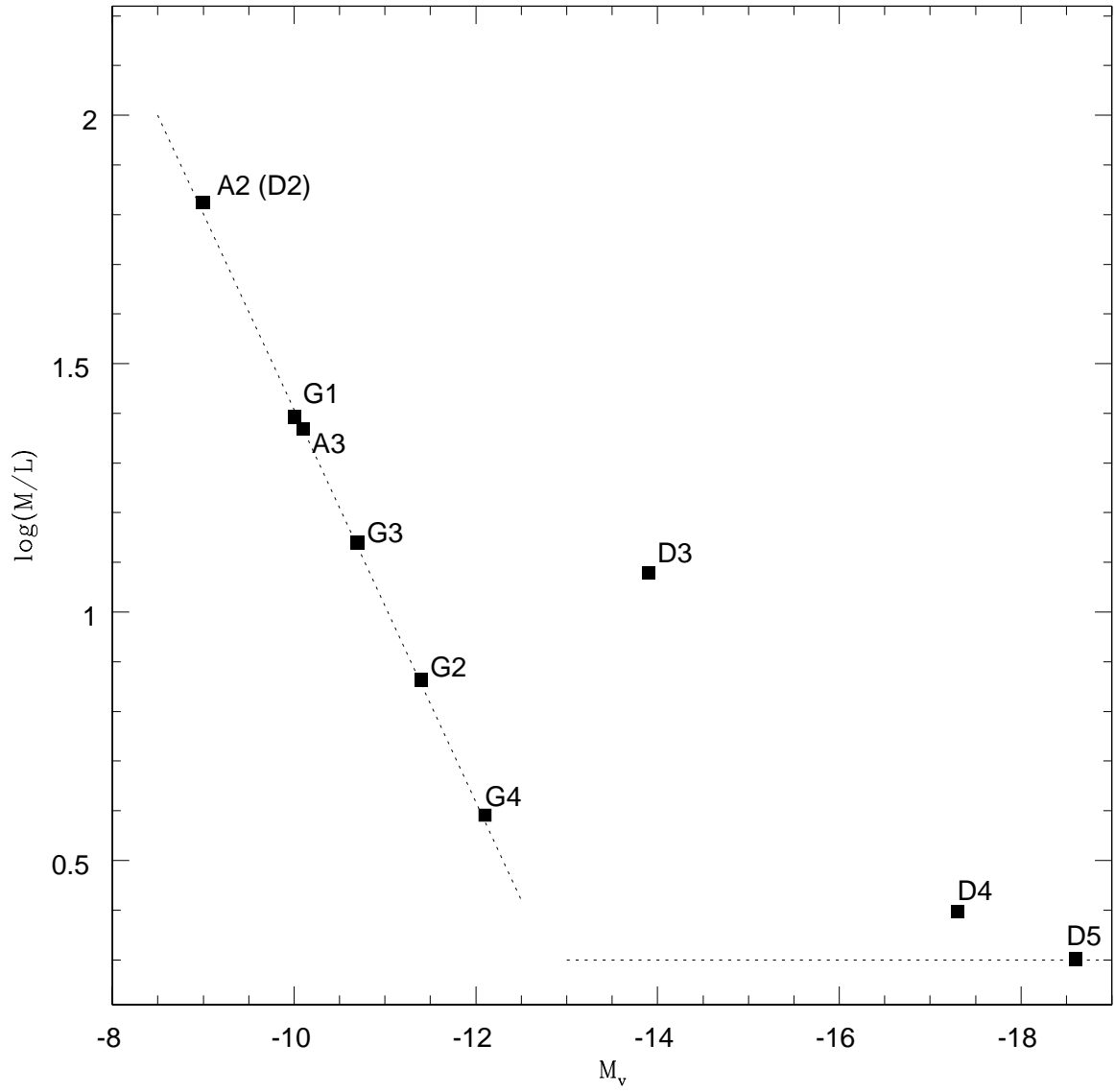}
\caption{Mass-to-light ratio versus $M_v$ for models A1-3, D1-5 and G0-5.}
\label{fig:mlmv}
\end{figure}

Observational data suggest that in the satellite dSphs around the
Galaxy, the baryonic matter content varies widely, while the dark
matter not only dominates but also has widely varying masses.  But, in
dwarf galaxies with mass more than $\sim 10^{7-8} M_\odot$, the total
mass-to-light ratio has a nearly constant value $\sim$ 2.0
(see Mateo {\it et al.} 1998).  The results of our calculation are
consistent with this
apparent dichotomy.  In order to make a direct comparison with Fig.~7
in Mateo {\it et al.} (1998), we plot the total mass to light ratio
$M/L$ within $r_0$ as a function of the absolute visual magnitude
$M_v$ of the different dwarf galaxies in models D1-5, A1-3, and G0-5.
For very low-mass ($M_0 <5 \times 10^6 M_\odot$) dark matter halo,
star formation may be completely suppressed even if $b$ is 0.5.  In
these low-mass systems, the pressure scale height of the photoionized
gas is comparable to the galaxy's radius and in the vicinity of the
Galactic disks, and the tidal perturbation of nearby host galaxy would
induce efficient mass loss (Murray {\it et al.} 2003).  It is entirely
possible that that the halo of the Milky Way may be populated by many
dark matter halos of the `missing dwarf galaxies'.  For the
intermediate-mass ($M_0 \sim 10^{7-8}$) dwarf galaxies, the gas
retention efficiency depends on many factors such as the background UV
flux, the IMF, the baryonic to dark matter ratio etc.
In Figure~\ref{fig:mlmv},
the $(M/L)$-$M_v$ correlation for this mass range merely indicate that a
transition mass for the gas retention efficiency.  For the high-mass
dwarf galaxies, the baryonic matter represents a significant fraction
of the total mass and the constant $M/L$ is a reflection of the
self-regulated star formation process.

\subsection{Multiple Bursts of Star Formation}

In higher mass systems, feedback occurs before the gas can evolve
significantly.  Star formation is therefore able to occur at an
almost equilibrium, self-regulated rate.

By contrast, in low mass systems, $\tau_{ff}+\tau_{onset}$ exceeds the
dynamical time on which the gas evolves within the galactic potential.
If sufficient gas is present to shield the core from external UV, then
gas in the core may accumulate significantly before the onset of feedback,
leading to strong bursts of star formation.  The same is
observed in larger systems, for which we have considered larger values
of $\tau_{onset}$.

If the initial burst of star formation causes the gas to expand, and its
density to decrease to the point at which the external UV is able to
ionize the galaxy completely, then star formation is terminated after a
single burst.  In somewhat more massive systems, gas in the core is still
self-shielded after the massive stars formed in the initial burst fade
away.  The gas can then cool, contract, and trigger additional bursts of
star formation, as is observed in some of our models.  In all systems
which undergo either one or multiple episodes of star formation, the
feedback is eventually sufficiently strong as to drive the gas out of the
system.

Episodic star formation may also result from reducing the efficiency
of feedback, either by reducing the conversion efficiency of gas into
stars, or the fraction of massive stars which form.  In high mass models
with such variations, we observe episodic star formation.

\subsection{Induced Star Formation}

Within the environment of a galaxy cluster, or near to a massive
galaxy, a dwarf system shall experience tidal forces, that shall vary
over time, resulting in occasional, or even periodic compressions.  If
dwarf systems form in large numbers around more massive overdense
regions, they shall also experience collisions.  The merger process
which builds larger galactic entities can also lead to collisions and
disturbances.

Both mechanisms lead to enhancements in the density, and therefore the
recombination efficiency of the gas within dwarf galaxies.  In small
systems at early epochs, this may allow the galaxy to form some stars,
when it might otherwise be completely ionized by the external UV
radiation.  In more massive systems, the star formation rates are
enhanced by as much as an order of magnitude relative to unperturbed
systems for realistic compression speeds.

Perturbations such as considered here would be expected to be stronger
in dense cluster environments, or near to massive galaxies.  In the
former, collisions would be more frequent, while in the latter,
tidal variations due to elliptical orbits around the parent system are
strongest.  These are also environments where tidal stripping, ram
pressure, and external UV radiation are all enhanced, which would be
expected to act to reduce star formation efficiencies, and so it is
unclear whether the effect of perturbations alone would lead to
environmental dependences for stellar populations within dwarf
galaxies.

\subsection{Consequences for Dwarf Galaxy Evolution}

As indicated in earlier work, star formation in the lowest mass dwarf
galaxies is difficult at early epochs, and the observed population of
dSphs appear to be at the very edge of the ability of small galaxies
to form stars.  The difficulty is due to the
strong UV background, which easily ionizes the galaxies throughout,
preventing the gas from cooling and contracting within the weak
gravitational potential of the dwarf system.

Star formation may still occur, even in the smallest systems, however, if
the gas within them can become self-shielded from the external UV.  Such
might be the case if they form with a relatively large baryonic fraction,
if they accrete additional gas, or if they undergo compressions due to
either tidal forces or collisions.  Such processes may help to explain
the existence of old stellar populations in even the smallest dSphs.
Feedback from star formation also forms a natural means for avoiding
the ``over-cooling'' problem among the dwarf progenitors
of massive galaxies.

In small systems, star formation tends to occur in bursts, the feedback
from which either expels the gas completely from the galaxy, or drives
it to large radii, making it vulnerable to either tidal or ram-pressure
stripping.  Subsequent generations of star formation would, therefore,
appear to be unlikely.  The gas lost to a dwarf system, however, leaves at
small speeds ($\sim10$~km~s$^{-1}$).  It shall not, therefore, move to
large distances from the dwarf as it orbits within the potential of either
a massive galaxy or within a cluster, and may be re-accreted by the dwarf
at apogalacticon, possibly leading to subsequent bursts of star formation.
Some small dwarf systems, therefore, might be expected to show periodic
bursts of
star formation, with a burst period determined by the orbital period of the
dwarf.

By contrast, in larger systems, where the feedback time is short compared
to the dynamical time of the system, and where the potential is capable
of retaining ionized gas, feedback tends to lead to steady, self-regulated
star formation.  The gas within the system does, however, expand under the
influence of stellar feedback, and so may also be subject to stripping
within galactic cluster environments (Murray, Dong, \& Lin 2003).

The possibility of severe mass loss from dwarf galaxies following star
formation indicates that the currently observed baryonic-to-dark matter
ratios can provide only lower limits, possibly very weak limits, to the
values present when the galaxies formed.  Our models indicate that the
star formation efficiencies, and the ability of galaxies to retain gas
following star formation depend very sensitively dependences upon the
evolutionary histories of the galaxies, and upon the galactic potentials,
which may account for the large dispersion in the stellar to dark-matter
fraction observed today among the dwarf galaxies.

\begin{acknowledgments}
We thank Drs. A. Burkert, and K-S. Oh for useful conversations.  We also
thank the anonymous referee for valuable comments. This work
was performed under the auspices of the U.S. Department of Energy by University
of California, Lawrence Livermore National Laboratory under Contract
W-7405-Eng-48.  This work is partially supported by NASA through an
astrophysical theory grant NAG5-12151.
\end{acknowledgments}

\clearpage

\begin{deluxetable}{cclrcrcr}
\tablewidth{0pt}
\tablecaption{Static Galaxy Models \label{tab:galaxymodels1}}
\tablehead{
\colhead{Model} &
\colhead{r$_0$} &
\colhead{$b$} &
\colhead{$\frac{J_\nu(z)}{J_\nu(0)}$} &
\colhead{$\beta_{max}$} &
\colhead{$f_{IMF}$} &
\colhead{$\tau_{onset}$} &
\colhead{$f_{\ast}$} \\
\colhead{} &
\colhead{(kpc)} &
\colhead{} &
\colhead{} &
\colhead{} &
\colhead{} &
\colhead{(Myr)} &
\colhead{(\%)}
}
\startdata
A1 & 0.6 & 0.001 & 1   & 1.0 &  1 & 0.5 &  0.0 \\
A2 & 0.6 & 0.1   & 1   & 1.0 &  1 & 0.5 &  2.6 \\
A3 & 0.6 & 0.2   & 1   & 1.0 &  1 & 0.5 &  3.7 \\
B1 & 0.3 & 0.1   & 1   & 1.0 &  1 & 0.5 &  1.3 \\
B2 & 0.3 & 0.2   & 1   & 1.0 &  1 & 0.5 &  1.9 \\
B3 & 0.3 & 0.5   & 1   & 1.0 &  1 & 0.5 &  2.5 \\
C1 & 0.3 & 0.1   & 100 & 1.0 &  1 & 0.5 &  0.0 \\
C2 & 0.3 & 0.2   & 100 & 1.0 &  1 & 0.5 &  0.7 \\
C3 & 0.3 & 0.5   & 100 & 1.0 &  1 & 0.5 &  1.5 \\
D1 & 0.3 & 0.1   & 1   & 1.0 &  1 & 0.5 &  1.3 \\
D2 & 0.6 & 0.1   & 1   & 1.0 &  1 & 0.5 &  2.6 \\
D3 & 2.0 & 0.1   & 1   & 1.0 &  1 & 0.5 & 14.2 \\
D4 & 4.0 & 0.1   & 1   & 1.0 &  1 & 0.5 & 68.4 \\
D5 & 6.0 & 0.1   & 1   & 1.0 &  1 & 0.5 & 83.2 \\
E1 & 0.3 & 0.1   & 100 & 1.0 &  1 & 0.5 &  0.0 \\
E2 & 0.6 & 0.1   & 100 & 1.0 &  1 & 0.5 &  0.0 \\
E3 & 2.0 & 0.1   & 100 & 1.0 &  1 & 0.5 & 19.7 \\
E4 & 4.0 & 0.1   & 100 & 1.0 &  1 & 0.5 & 66.3 \\
E5 & 6.0 & 0.1   & 100 & 1.0 &  1 & 0.5 & 67.6 \\
F1 & 4.0 & 0.01  & 0   & 1.0 &  1 & 0.5 & 24.5 \\
F2 & 4.0 & 0.01  & 1   & 1.0 &  1 & 0.5 & 24.1 \\
F3 & 6.0 & 0.01  & 0   & 1.0 &  1 & 0.5 & 47.7 \\
F4 & 6.0 & 0.01  & 1   & 1.0 &  1 & 0.5 & 45.4 \\
G0 & 0.6 & 0.2   & 1   & 1.0 &  1 & 0.5 &  3.7 \\
G1 & 0.6 & 0.2   & 1   & 0.1 &  1 & 0.5 &  3.5 \\
G2 & 0.6 & 0.2   & 1   & 1.0 & 10 & 0.5 & 11.7 \\
G3 & 0.6 & 0.2   & 1   & 1.0 &  1 & 5.0 &  6.3 \\
G4 & 0.6 & 0.2   & 1   & 0.1 & 10 & 5.0 & 22.4 \\
H0 & 4.0 & 0.1   & 1   & 1.0 &  1 & 0.5 & 68.4 \\
H1 & 4.0 & 0.1   & 1   & 1.0 &  1 & 5.0 & 71.7 \\
H2 & 4.0 & 0.1   & 1   & 1.0 &  1 & 50. & 35.0 \\
H3 & 4.0 & 0.1   & 1   & 0.1 &  1 & 5.0 & 69.7 \\
H4 & 4.0 & 0.1   & 1   & 0.01&  1 & 5.0 & 49.4 \\
H5 & 4.0 & 0.1   & 1   & 0.1 & 10 & 5.0 & 85.7 \\
\enddata
\end{deluxetable}

\begin{deluxetable}{cclrcrcr}
\tablewidth{0pt}
\tablecaption{Externally Perturbed Galaxy Models \label{tab:galaxymodels2}}
\tablehead{
\colhead{Model} &
\colhead{r$_0$} &
\colhead{$b$} &
\colhead{$\frac{J_\nu(z)}{J_\nu(0)}$} &
\colhead{$H_p/H_0$} &
\colhead{$f_{IMF}$} &
\colhead{$\tau_{onset}$} &
\colhead{$f_{\ast}$} \\
\colhead{} &
\colhead{(kpc)} &
\colhead{} &
\colhead{} &
\colhead{} &
\colhead{} &
\colhead{(Myr)} &
\colhead{(\%)}
}
\startdata
I0 & 0.6 & 0.1   & 1   & 0     &  1 & 0.5  &  3.7 \\
I1 & 0.6 & 0.1   & 1   & 3     &  1 & 0.5  &  4.2 \\
I2 & 0.6 & 0.1   & 1   & 15    &  1 & 0.5  &  6.3 \\
I3 & 0.6 & 0.1   & 1   & 30    &  1 & 0.5  & 27.3 \\
I4 & 0.6 & 0.1   & 1   & 150   &  1 & 0.5  & 89.3 \\
K0 & 2.0 & 0.1   & 1   & 0     &  1 & 0.5  & 14.2 \\
K1 & 2.0 & 0.1   & 1   & 3     &  1 & 0.5  & 32.1 \\
K2 & 2.0 & 0.1   & 100 & 0     &  1 & 0.5  & 19.7 \\
K3 & 2.0 & 0.1   & 100 & 3     &  1 & 0.5  & 23.7 \\
\enddata
\end{deluxetable}

\clearpage


\clearpage

\clearpage

\clearpage

\clearpage

\clearpage

\clearpage

\clearpage

\clearpage

\clearpage

\clearpage

\clearpage

\clearpage

\clearpage

\end{document}